\documentstyle[11pt,psfig]{article} 
\mathcode`\0="0030      %
\mathcode`\1="0031
\mathcode`\2="0032
\mathcode`\3="0033
\mathcode`\4="0034
\mathcode`\5="0035
\mathcode`\6="0036
\mathcode`\7="0037
\mathcode`\8="0038
\mathcode`\9="0039

\makeatletter
\def\@begintheorem#1#2{\trivlist\item[\hskip\labelsep{\bf #1\ #2}]}
\def\foobarpt{\textfont\z@\tenrm 
  \scriptfont\z@\ninrm \scriptscriptfont\z@\sevrm
\textfont\@ne\tenmi \scriptfont\@ne\ninmi \scriptscriptfont\@ne\sevmi
\textfont\tw@\tensy \scriptfont\tw@\ninsy \scriptscriptfont\tw@\sevsy
\textfont\thr@@\tenex \scriptfont\thr@@\tenex \scriptscriptfont\thr@@\tenex
\def\unboldmath{\everymath{}\everydisplay{}\@nomath\unboldmath
          \textfont\@ne\tenmi 
          \textfont\tw@\tensy \textfont\lyfam\tenly
          \@boldfalse}\@boldfalse
\def\boldmath{\@ifundefined{tenmib}{\global\font\tenmib\@mbi\@magscale1\global
        \font\tensyb\@mbsy \@magscale1\global\font
         \tenlyb\@lasyb\@magscale1\relax\@addfontinfo\@xiipt
              {\def\boldmath{\everymath
                {\mit}\everydisplay{\mit}\@prtct\@nomathbold
                \textfont\@ne\tenmib \textfont\tw@\tensyb 
                \textfont\lyfam\tenlyb\@prtct\@boldtrue}}}{}\@xiipt\boldmath}%
\def\prm{\fam\z@\tenrm}%
\def\pit{\fam\itfam\tenit}\textfont\itfam\tenit \scriptfont\itfam\ninit
   \scriptscriptfont\itfam\sevit
\def\psl{\fam\slfam\tensl}\textfont\slfam\tensl 
     \scriptfont\slfam\tensl \scriptscriptfont\slfam\tensl
\def\pbf{\fam\bffam\tenbf}\textfont\bffam\tenbf 
   \scriptfont\bffam\ninbf \scriptscriptfont\bffam\ninbf 
\def\ptt{\fam\ttfam\tentt}\textfont\ttfam\tentt
   \scriptfont\ttfam\nintt \scriptscriptfont\ttfam\nintt 
\def\psf{\fam\sffam\tensf}\textfont\sffam\tensf
    \scriptfont\sffam\tensf \scriptscriptfont\sffam\tensf
\def\psc{\@getfont\psc\scfam\@xiipt{\@mcsc\@magscale1}}%
\def\ly{\fam\lyfam\tenly}\textfont\lyfam\tenly 
   \scriptfont\lyfam\ninly \scriptscriptfont\lyfam\sevly
 \@setstrut \rm}

\newcommand{\singlespacing}{\let\CS=
\@currsize\renewcommand{\baselinestretch}{1}\tiny\CS}
\newcommand{\singlespacingplus}{\let\CS=
\@currsize\renewcommand{\baselinestretch}{1.25}\tiny\CS}
\newcommand{\doublespacing}{\let\CS=
\@currsize\renewcommand{\baselinestretch}{1.75}\tiny\CS}
\newcommand{\draftspacing}{\let\CS=
\@currsize\renewcommand{\baselinestretch}{2.0}\tiny\CS}

\newcommand{\niceonespacing}{\let\CS=\@currsize\renewcommand{\baselinestretch}{1.1}\tiny\CS}
\newcommand{\nicetwospacing}{\let\CS=\@currsize\renewcommand{\baselinestretch}{1.2}\tiny\CS}
\newcommand{\nicethreespacing}{\let\CS=\@currsize\renewcommand{\baselinestretch}{1.3}\tiny\CS}
\newcommand{\singlespacingplusplus}{\let\CS=\@currsize\renewcommand{\baselinestretch}{1.35}\tiny\CS}
\newcommand{\nicefivespacing}{\let\CS=\@currsize\renewcommand{\baselinestretch}{1.5}\tiny\CS}
\newcommand{\nicesixpacing}{\let\CS=\@currsize\renewcommand{\baselinestretch}{1.6}\tiny\CS}
\newcommand{\nicefoospacing}{\let\CS=\@currsize\renewcommand{\baselinestretch}{1.15}\tiny\CS}

\makeatother

\makeatletter
\def\@cite#1#2{[#1\if@tempswa , #2\fi]}
\makeatother

\newcommand\seq{\subseteq}
\renewcommand\.{\cdot}
\newcommand\<{\langle}
\renewcommand\>{\rangle}

\newcommand\Lora{\ \Longrightarrow \ }
\newcommand\Lolra{\ \Longleftrightarrow \ }

\newcommand{\sigmastar}{\mbox{$\Sigma^\ast$}}
\newcommand\equalsdef{\stackrel{\mbox{\protect\scriptsize df}}{=}}
\newcommand\tweak{\hspace*{1pt}}

\newcommand\Oplus{\mbox{\boldmath $\oplus$}}
\newcommand\xor{\mbox{$\sym$}}
\newcommand\XOR{\mbox{\boldmath $\sym$}}
\newcommand\nxor{\mbox{$\overline{\sym}$}}
\newcommand\NXOR{\mbox{\boldmath $\overline{\sym}$}}
\newcommand\union{\mbox{\boldmath $\vee$}}
\newcommand\inter{\mbox{\boldmath $\wedge$}}

\def\sym{\Delta}

\newcommand\N{{\rm I\!N}}

\newcommand\p{\mbox{\rm P}}
\newcommand\fp{\mbox{\rm FP}}
\newcommand\np{\mbox{\rm NP}}

\newcommand\co{\mbox{\rm co}}

\newcommand{\psel}{\mbox{\rm P-Sel}}
\newcommand{\ppoly}{\mbox{\rm P/poly}}

\newtheorem{theorem}{Theorem}[section]
\newtheorem{corollary}[theorem]{Corollary}
\newtheorem{lemma}[theorem]{Lemma}

\newtheorem{remark}[theorem]{Remark}

\newtheorem{proposition}[theorem]{Proposition}
\newtheorem{definition}[theorem]{Definition}

\newenvironment{construction}{\bigbreak\begin{block}}{\end{block}
    \bigbreak}

\newenvironment{block}{\begin{list}{\hbox{}}{\leftmargin 1em
    \itemindent -1em \topsep 0pt \itemsep 0pt \partopsep 0pt}}{\end{list}}

\begin{document}

\title{Polynomial-Time Multi-Selectivity}

\author{
{\em  Lane A. Hemaspaandra\/} 
\protect\thanks{
Department of Computer Science,
University of Rochester,
Rochester, NY 14627, USA. 
}
\and
{\em  Zhigen Jiang\/} 
\protect\thanks{
Institute of Software,
Chinese Academy of Sciences,
Beijing 100080, China. 
}
\and
{\em  J\"org Rothe\/} 
\protect\thanks{
Institut f\"ur Informatik,
Friedrich-Schiller-Universit\"at Jena,
07743 Jena, Germany.
}
\and
{\em  Osamu Watanabe\/} 
\protect\thanks{
Department of Computer Science,
Tokyo Institute of Technology,
Tokyo 152, Japan.
}
} %

\newcount\hour  \newcount\minutes  \hour=\time  \divide\hour by 60
\minutes=\hour  \multiply\minutes by -60  \advance\minutes by \time
\def\mmmddyyyy{\ifcase\month\or Jan\or Feb\or Mar\or Apr\or May\or Jun\or Jul\or
  Aug\or Sep\or Oct\or Nov\or Dec\fi \space\number\day, \number\year}
\def\hhmm{\ifnum\hour<10 0\fi\number\hour :%
  \ifnum\minutes<10 0\fi\number\minutes}
\def\Draft{{\it Draft of \mmmddyyyy}}

\date{}

\typeout{WARNING:  BADNESS used to supress reporting!  Beware!!}
\hbadness=3000%
\vbadness=10000 %

\setcounter{footnote}{0}
{\singlespacing
{\maketitle}
}

\begin{abstract}
We introduce a generalization of Selman's
P-selectivity that yields a more
flexible notion of selectivity, called (polynomial-time)
multi-selectivity, in which the selector is allowed to operate on
multiple input strings.  Since our introduction of 
this class, it has been used~\cite{hem-jia-rot-wat:c:multiselectivity}
to prove the first known
(and optimal) lower bounds for generalized selectivity-like classes in
terms of $\mbox{\rm EL}_2$, the second level of the extended low
hierarchy. We study the resulting selectivity
hierarchy, denoted by SH, which we prove does not collapse. In
particular, we study the internal structure and the properties of SH
and completely establish, in terms of incomparability and strict
inclusion, the relations between our generalized selectivity classes
and Ogihara's P-mc (polynomial-time membership-comparable) classes.
Although SH is a strictly increasing infinite hierarchy, we show that
the core results that hold for the P-selective sets and that prove them
structurally simple also hold for SH\@.  In particular, all sets in
SH have small circuits; the NP sets in SH are in~$\mbox{\rm Low}_2$, the
second level of the low hierarchy within NP; and SAT
cannot be in SH unless $\p = \np$\@.  Finally, it is known that 
P-Sel, the
class of P-selective sets, is not closed under union or
intersection.  We provide an
extended selectivity hierarchy that is based on SH and that is large enough
to capture those closures of the P-selective sets, and yet, in
contrast with the P-mc classes, is refined enough to distinguish them.
\end{abstract}

\section{Introduction}

Selman introduced the P-selective sets (\psel, for short)
\cite{sel:j:pselective-tally} as the complexity-theoretic analogs of
Jockusch's semi-recursive sets~\cite{joc:j:semi}: A set is P-selective
if there exists a polynomial-time transducer (henceforward called a
selector) that, given any two input strings, outputs one that is
logically no less likely to be in the set than the other one.  There
has been much progress recently in the study of P-selective sets (see
the survey \cite{den-han-hem-tor:j:semi-membership}).  In this paper,
we introduce a more flexible notion of selectivity that allows the
selector to operate on multiple input strings, and that thus
generalizes Selman's P-selectivity in the following promise-like way:
Depending on two parameters, say $i$ and $j$ with $i\geq j\geq 1$, a
set $L$ is $(i,j)$-selective if there is a selector that, given any
finite set of distinct input strings, outputs some subset of at least
$j$ elements each belonging to $L$ if $L$ contains at least $i$ of the
input strings; otherwise, it may output an arbitrary subset of the
inputs.  Observe that in this definition of $(i,j)$-selectivity only
the difference of $i$ and $j$ is relevant: $L$ is $(i,j)$-selective if
and only if $L$ is $(i-j+1,1)$-selective.  Let $\mbox{\rm S}(k)$ denote
the class of $(k,1)$-selective sets.  Clearly, $\mbox{\rm S}(1) = \psel$,
and for each $k\geq 1$, $\mbox{\rm S}(k) \seq \mbox{\rm S}(k+1)$. This paper
is devoted to the study of the resulting hierarchy, $\mbox{\rm SH}
\equalsdef \bigcup_{k \geq 1} \mbox{\rm S}(k)$.

The literature contains many notions that
generalize
P-selectivity.  For example, 
Ko's ``weakly P-selective
sets''~\cite{ko:j:self-reducibility}, Amir, Beigel, and Gasarch's
``non-p-superterse sets''~\cite{ami-bei-gas:conf-plus-manu:uni}
(sometimes called ``approximable
sets''~\cite{bei-kum-ste:j:approximable-sets}), Ogihara's
``polynomial-time membership-comparable
sets''~\cite{ogi:j:comparable}, Cai and Hemaspaandra's (then
Hemachandra) ``polynomial-time enumerable sets''
(\cite{cai-hem:j:enum}, see the discussion
in~\cite{ogi:j:comparable}), and the ``${\cal FC}$-selective sets for
arbitrary function classes~${\cal FC}$'' of Hemaspaandra et
al.~\cite{hem-hoe-nai-ogi-sel-thi-wan:j:np-selective} 
all are notions generalizing P-selectivity.

Given the number of already known and well-studied generalizations of
P-Sel, the first question that naturally arises is: Why should one 
introduce another generalization of P-Sel? One motivation comes
from other results of this paper's 
authors (\cite{hem-jia-rot-wat:c:multiselectivity}, see
also~\cite{rot:thesis:promise}), which---in terms of the selectivity 
notion proposed in this paper---establish the first known (and optimal)
lower bounds for generalized selectivity-like classes 
with regard to~$\mbox{\rm EL}_2$, the second level of the extended low
hierarchy~\cite{bal-boo-sch:j:low}. 
In particular,
there exists
a sparse set in $\mbox{\rm S}(2)$ that is not 
in~$\mbox{\rm EL}_2$~\cite{hem-jia-rot-wat:c:multiselectivity,rot:thesis:promise}.
This
sharply contrasts with the known result that all P-selective sets are
in~$\mbox{\rm EL}_2$. The proof of this $\mbox{\rm EL}_2$ lower bound
additionally creates another interesting result: $\mbox{\rm EL}_2$ is not
closed under certain Boolean connectives such as union and intersection.  
This extends the known result that P-Sel is not closed under 
those Boolean connectives~\cite{hem-jia:j:psel}.
Finally, the proof technique used to show the $\mbox{\rm EL}_2$ lower
bounds for generalized selectivity classes can be adapted to give the
main result of~\cite{hem-jia-rot-wat:c:multiselectivity}: 
There exist sets that are not in
$\mbox{EL}_2$, yet their join is in~$\mbox{EL}_2$.  
That is, the join operator can lower
difficulty as measured in terms of extended lowness. 
Since in a strong intuitive sense the
join does not lower complexity, this result suggests that, if one's intuition
about complexity is---as is natural---based on reductions,
then the extended low hierarchy is not a natural measure
of complexity.  Rather, it is a measure that is related
to the difficulty of information extraction, and it is 
in flavor quite orthogonal to more traditional notions
of complexity. 

Another motivation for the study of the multi-selective sets is closely
related to the known results mentioned in the previous paragraph.
Since P-Sel is not closed under union or intersection, it is natural
to ask which complexity classes are appropriate to capture, e.g., the
class of intersections of P-selective sets. Even more to the point, can the
intersections (or the unions) of P-selective sets be {\em
  classified\/} in some complexity-theoretic setting, for instance by
proving that the class of intersections of P-selective sets is
contained in such-and-such level of some hierarchy of complexity
classes, but not in the immediately
lower level?  Though we will show that SH
is not appropriate to provide answers to questions like this (since we
prove that the above-mentioned result on unions and intersections
extends to all levels of SH,
i.e., neither the closure of P-Sel under union nor the closure of 
P-Sel under intersection is contained in any level of~SH), 
we will introduce in Section~\ref{sec:gch} an
extended selectivity hierarchy that is based on SH and can be used to
classify Boolean closures of P-selective sets.

This paper is organized as follows. In Section~\ref{notations}, we provide
our notations and some definitions.
In Section~\ref{subsec:p-mc}, we
study the internal structure and the properties of~SH. In particular,
we show that SH is properly infinite, and we relatedly prove that,
unlike P-Sel, none of the $\mbox{\rm S}(k)$ for $k\geq 2$ is closed under
$\leq^p_m$-reductions, and also that sets in $\mbox{\rm S}(2)$ that are
many-one reducible to their complements may already go beyond~P,
which contrasts with Selman's result that a set $A$ is in \p\ if and
only if $A \leq_{m}^{p} \overline{A}$ and $A$ is 
P-selective~\cite{sel:j:pselective-tally}. 
Consequently, the class P cannot be
characterized by the auto-reducible sets in any of the higher levels
of~SH\@.  This should be compared with Buhrman and
Torenvliet's nice characterization of P as those self-reducible sets
that are in P-Sel~\cite{buh-tor:j:pselective}.

We then compare the levels of SH with the levels of Ogihara's
hierarchy of polynomial-time membership-comparable (P-mc, for short)
sets. Since $\mbox{\rm P-mc}(k)$ (see Definition~\ref{def:p-mc}) is closed
under $\leq_{1\mbox{-}tt}^{p}$-reductions for
each~$k$~\cite{ogi:j:comparable}, it is clear from the provable
non-closure under $\leq^p_m$-reductions of the $\mbox{\rm S}(k)$, $k\geq
2$, that Ogihara's approach to generalized selectivity is different
from ours, and in Theorem~\ref{thm:fair-s-p-mc}, we completely
establish, in terms of incomparability and strict inclusion, the
relations between his and our generalized selectivity classes. In
particular, since \mbox{$\mbox{\rm P-mc}(\mbox{poly})$} is contained in
$\ppoly$~\cite{ogi:j:comparable} and SH is (strictly) contained in
$\mbox{\rm P-mc}(\mbox{poly})$, it follows that every set in SH has
polynomial-size circuits. On the other hand, P-selective NP sets can
even be shown to be in $\mbox{\rm Low}_2$~\cite{ko-sch:j:circuit-low}.
Since such a result is not known to hold for the polynomial-time
membership-comparable NP sets, our $\mbox{\rm Low}_2$-ness results in
Theorem~\ref{thm:lowness} are the
strongest known for generalized selectivity-like classes.
(Note, however, that K\"obler~\cite{koe:c:low-survey} has observed
that our generalization of Ko and Sch\"oning's result that
\mbox{$\psel \cap \np \seq \mbox{\rm
    Low}_2$}~\cite{ko-sch:j:circuit-low} can be combined with other
generalizations of the same result to yield a very generalized
statement, as will be explained in more detail near the start of
Section~\ref{subsec:circuit}.)

Selman proved that NP-complete sets such as SAT (the satisfiability
problem) cannot be P-selective unless \mbox{$\p =
  \np$}~\cite{sel:j:pselective-tally}.  Ogihara extended this collapse
result to the case of certain P-mc classes strictly larger
than~\psel. 
By the inclusions stated in Theorem~\ref{thm:fair-s-p-mc}, this extension 
applies to many of our selectivity classes as well; in particular, SH cannot 
contain all of NP unless $\p = \np$.

To summarize, the results claimed in the previous two paragraphs
(and to be proven in Section~\ref{subsec:circuit})
demonstrate that the core results holding for the P-selective sets
and proving them structurally simple also hold for~SH.

In Section~\ref{subsec:capture}, we show into which levels of Ogihara's P-mc
hierarchy the closures of P-Sel under certain Boolean operations fall.
In particular, we prove that the closure of P-Sel under union and the
closure of P-Sel under intersection fall into exactly the same level
of the P-mc hierarchy and are not contained in the immediately lower level,
which shows they are indistinguishable in terms of \mbox{P-mc} classes.  
We also show that the closure of P-Sel under certain Boolean
operations is not contained in any level of~SH.
We then provide an extended selectivity hierarchy that is based on SH and
is large enough to capture those closures of P-selective sets, and
yet, in contrast with the P-mc classes, is refined enough to
distinguish them.  Finally, we study the internal structure of this
extended selectivity hierarchy in Section~\ref{subsec:structure}. 
The proofs of some of the more technical results in 
Section~\ref{subsec:structure} are deferred to 
Section~\ref{subsec:hard-proofs}.

\section{Notations and Definitions}
\label{notations}

In general, we adopt the standard notations of Hopcroft and
Ullman~\cite{hop-ull:b:automata}.  We consider sets of strings over
the alphabet $\Sigma\equalsdef \{0,1\}$.  For each string $x\in \Sigma^{*}$,
$|x|$ denotes the length of $x$.  For $k\geq 1$, let $x^k \equalsdef  x\cdot
x^{k-1}$, where $x^0 \equalsdef  \epsilon$ is the empty string and the dot
denotes the concatenation of strings.  ${\cal P}(\Sigma^{*})$ is the 
class of sets of strings over $\Sigma$.  
Let $\N$ (respectively, $\N^{+}$) denote the set of non-negative 
(respectively, positive) integers.
For any set $L\subseteq \Sigma^{*}$, 
$\|L\|$ represents the cardinality of $L$, and
$\overline{L}\equalsdef \Sigma^{*}-L$ denotes the complement of $L$ in
$\Sigma^{*}$. 

For sets $A$ and $B$, their {\em join\/}, 
$A\oplus B$\label{ind:join}, is $\{0x\,|\,x\in
A\}\cup\{1x\,|\,x\in B\}$, and the Boolean operations\label{ind:boolean} 
{\em symmetric difference\/}
(also called exclusive-or) and
{\em equivalence\/} (also called nxor) are defined as 
\mbox{$A \xor B\label{ind:xor} \equalsdef (A
\cap \overline{B}) \cup (\overline{A} \cap B)$} and 
\mbox{$A \nxor B\label{ind:nxor}
\equalsdef (A \cap B) \cup (\overline{A} \cap \overline{B})$}.  For
any class~${\cal C}$, define \mbox{$\co \tweak{\cal C}
\equalsdef \{ L \,|\, \overline{L} \in {\cal C}\}$}.
For classes $\cal C$ and $\cal D$ of
sets, define
\[
\begin{array}{lcllcl}
{\cal C} \,\inter\,\label{ind:inter} {\cal D} & \equalsdef & 
\{A \cap B \,|\, A\in {\cal C}\wedge B\in {\cal D}\}, &
{\cal C} \,\XOR\,\label{ind:XOR} {\cal D}   & \equalsdef &
\{A \,\xor\, B \,|\, A\in {\cal C}\wedge B\in {\cal D}\}, \\
{\cal C} \,\union\,\label{ind:union} {\cal D} & \equalsdef & 
\{A \cup B \,|\, A\in {\cal C}\wedge B\in {\cal D}\}, &
{\cal C} \,\NXOR\,\label{ind:NXOR} {\cal D}  & \equalsdef & 
\{A \,\nxor\, B \,|\, A\in {\cal C}\wedge B\in {\cal D}\}, \\
{\cal C} \,\Oplus\,\label{ind:Join} {\cal D} & \equalsdef & 
\{A \oplus B\,|\,A\in {\cal C}\wedge B\in {\cal D}\}. 
\end{array}
\]
For $k$ sets $A_1 , \ldots , A_k$, the join extends to 
$$
\oplus_k(A_1, \ldots , A_k) \equalsdef \bigcup_{1 \leq i \leq k} \{
\underline{i}x \mid x\in A_i \}, 
$$
where $\underline{i}$ is the bit
pattern of $\lceil \log k \rceil$ bits representing $i$ in binary.
We write $\Oplus_k({\cal C})$ to denote the class \mbox{$\{ 
\oplus_k(A_1,
  \ldots , A_k) \,|\, (\forall i : 1 \leq i \leq k)\, [A_i \in
  {\cal C}] \}$} of $k$-ary joins of sets in~${\cal C}$.  Similarly,
we use the shorthands $\inter_k({\cal C})$ and $\union_k({\cal C})$ to
denote the $k$-ary intersections and unions of sets in~${\cal C}$.

$L^{=n}$ (respectively, $L^{\leq n}$) is the set of strings in 
$L$ having length
$n$ (respectively, less than or equal to~$n$).  
Let \mbox{$\Sigma^n \equalsdef
  (\Sigma^*)^{=n}$}.  For a set $L$, $\chi_L$ denotes the
characteristic function of~$L$. The census function of $L$ is
defined by \mbox{$\mbox{\em census}_{L}(0^n) \equalsdef \| L^{\leq n} \|$}.
$L$ is said to be sparse if there is a polynomial $d$ such that for
any~$n$, \mbox{$\mbox{\em census}_{L}(0^n) \leq d(n)$}. Let SPARSE denote the
class of sparse sets.  To encode a pair of strings, we use a
polynomial-time computable pairing function, $\<\.,\.\>:\Sigma^*\times
\Sigma^*\rightarrow \Sigma^*$, that has polynomial-time computable
inverses; this notion is extended to encode every $m$-tuple of
strings, in the standard way.  Using the standard correspondence
between $\Sigma^*$ and $\N$, we will view $\<\.,\.\>$ also as a
pairing function mapping \mbox{$\N \times \N$} onto~$\N$\@.  A
polynomial-time transducer is a deterministic polynomial-time Turing
machine that computes functions from $\Sigma^{*}$ into $\Sigma^{*}$
rather than accepting sets of strings. FP denotes the class of
functions computed by polynomial-time transducers. Each selector
function considered is computed by a polynomial-time transducer that
takes a set of strings as input and outputs some set of strings.  As
the order of the strings in these sets doesn't matter, we may assume
that, without loss of generality, they are given in lexicographical
order (i.e., $x_1 \leq_{\mbox{\scriptsize lex}} x_2
\leq_{\mbox{\scriptsize lex}} \cdots \leq_{\mbox{\scriptsize lex}}
x_m$), and are coded into one string over $\Sigma$ using the above
pairing function. As a notational convenience, we'll identify these
sets with their codings and simply write (unless a more complete
notation is needed) $f(x_1,\ldots , x_m)$ to indicate that selector
$f$ runs on the inputs $x_1,\ldots , x_m$ coded as $\<x_1,\ldots ,
x_m\>$.

We shall use the shorthands NPM (NPOM) to refer to ``nondeterministic
polynomial-time (oracle) Turing machine.''  For an (oracle) Turing
machine $M$ (and an oracle set~$A$), $L(M)$ ($L(M^A)$) denotes the set
of strings accepted by $M$ (relative to~$A$). For any polynomial-time
reducibility $\leq^p_r$ and any class of sets ${\cal C}$, define
\mbox{${\Re}^{p}_{r} ({\cal C}) \equalsdef \{ L \mid (\exists C \in
{\cal C})\, [ L \leq^p_r C ] \}$}.  As is standard, E will denote
\mbox{$\bigcup_{c\ge 0}$DTIME$[2^{cn}]$}.

\begin{definition}\label{def:poly}
\cite{kar-lip:c:nonuniform}\quad
\ppoly\ denotes
the class of sets $L$ for which there exist a set $A\in \p$
and a polynomially length-bounded function
$h:\Sigma^* \rightarrow \Sigma^*$ such that for every~$x$,
{} it holds that
$x\in L$ if and only if $\<x,h(0^{|x|})\> \in A$.
\end{definition}

\begin{definition}\label{def:low}
\begin{enumerate}
\item {\cite{sch:j:low}}\quad 
For $k\geq 1$, define Low$_k \equalsdef \{ L\in \np \mid
\Sigma_k^{p,\,L} = \Sigma_k^p\}$, where the $\Sigma_k^p$
are the $\Sigma$ levels of the polynomial 
hierarchy~\cite{mey-sto:c:reg-exp-needs-exp-space,sto:j:poly}.

\item {\cite{bal-boo-sch:j:low,lon-she:j:low}}\quad
For $k\geq 2$, define EL$_k \equalsdef \{ L \mid
\Sigma_k^{p,\,L} =
\Sigma_{k-1}^{p,\,\mbox{\scriptsize SAT}\oplus L}\}$.
For $k\geq 3$, define EL$\Theta_k \equalsdef \{ L \mid
\p^{(\Sigma_{k-1}^{p,\,L})[\log n]} \subseteq
\p^{(\Sigma_{k-2}^{p,\,\mbox{\tiny SAT}\oplus L})[\log n]}\}$.
The $[\log n]$ indicates that at most ${\cal O}(\log n)$ queries
are made to the oracle.
\end{enumerate}
\end{definition}

\section{A Basic Hierarchy of Generalized Selectivity Classes}
\label{sec:structure}

\subsection{Structure, Properties, and Relationships with P-mc Classes}
\label{subsec:p-mc}

\begin{definition}
\label{def:S-fair-S}
\quad
Let $g_1$ and $g_2$ be non-decreasing functions from $\N^{+}$ into $\N^{+}$
(henceforward called %
{\em threshold functions\/})\label{ind:threshold} such that $g_1 \geq
g_2$.  $\mbox{\rm S}(g_1(\cdot),g_2(\cdot))$ is the class of sets $L$ for
which there exists an \fp\ function $f$ such that for each $n\geq 1$
and any distinct input strings $y_1,\ldots ,y_n$,
\begin{enumerate}
\item
$f(y_1,\ldots ,y_n) \seq \{y_1,\ldots ,y_n\}$, and

\item
if $\| L\cap \{y_1,\ldots ,y_n\}\| 
\geq g_1(n)$, then it holds that 
\mbox{$f(y_1,\ldots ,y_n) \seq L$} and
\mbox{$\| f(y_1,\ldots ,y_n)\| \geq g_2(n)$}.
\end{enumerate}
We also consider classes
$\mbox{\rm Fair-S}(g_1(\cdot),g_2(\cdot))$\label{ind:fair} 
in which the selector
$f$ is required to satisfy the above conditions only when applied to
any $n$ distinct input strings each having length at most~$n$. We will
refer to selectors having this property as selectors meeting the
``fairness condition.'' 
\end{definition}

As a notational convention and as a shorthand
for describing functions, for {\em non\/}-constant threshold
functions, we will use ``expressions in~$n$'' and we use~$i$,~$j$, or
$k$ if the threshold is constant.  
For example, rather than writing
${\rm S}(\lambda n.n-1\,,\,\lambda n.k)$, we will use the shorthand
${\rm S}(n-1,k)$, and rather than 
writing
$\mbox{\rm{}S}(\lambda n . g_1(n)\, , \, \lambda n . g_2(n))$
we will write
$\mbox{\rm{}S}(g_1(n),g_2(n))$.

Definition~\ref{def:S-fair-S} immediately implies
the following:

\begin{proposition}
\label{fact:basic}\quad
Let $g_1$, $g_2$, and $c$ be threshold functions such that $g_1 \geq g_2$. 
\begin{enumerate}
\item
\begin{enumerate}
\item 
$\mbox{\rm S}(g_1(n),g_2(n))
\seq \mbox{\rm S}(g_1(n)+c(n),g_2(n))$, and
\item 
$\mbox{\rm S}(g_1(n),g_2(n)+c(n)) \seq \mbox{\rm S}(g_1(n),g_2(n))$. 
\end{enumerate}
The above inclusions also hold for the corresponding Fair-S classes.

\item 
If $(\forall m) [g_1(m)\geq m]$,
then $\mbox{\rm S}(g_1(n),g_2(n)) = \mbox{\rm Fair-S}(g_1(n),g_2(n))
 = {\cal P}(\sigmastar)$.

\item
If $(\forall m)[g_2(m) \leq g_1(m) < m]$,
then $\mbox{\rm S}(g_1(n),g_2(n)) \seq \mbox{\rm Fair-S}(g_1(n),g_2(n))
\seq \mbox{\rm Fair-S}(n-1,1)$.
\end{enumerate}
\end{proposition}

In particular, we are interested in classes $\mbox{\rm S}(i,j)$\label{ind:sij}
parameterized by constants $i$ and $j$.
Theorem~\ref{thm:S-collapse} reveals that, in fact, there is only {\em
  one\/} significant parameter, the difference of $i$ and $j$.  This
suggests the simpler notation $\mbox{\rm S}(k)  \equalsdef  \mbox{\rm S}(k,1)$ for all
$k\geq 1$.  Let SH\label{ind:sh} denote the hierarchy $\bigcup_{k\geq 1}
\mbox{\rm S}(k)$.  For simplicity, we henceforward (i.e., after the proof
of Theorem~\ref{thm:S-collapse}) assume that selectors for any set in
SH select {\em exactly one\/} 
input string rather than a subset of the inputs (i.e.,
they are viewed as \fp\ functions mapping into $\sigmastar$ rather than
into~${\cal P}(\sigmastar)$).

\begin{theorem}\label{thm:S-collapse}\quad
$(\forall i\geq 1)\, (\forall k\geq 0)\, [\mbox{\rm S}(i,1)=\mbox{\rm S}(i+k,1+k)]$.
\end{theorem}

\noindent
{\bf Proof.} \quad
For any fixed $i\geq 1$, the proof is done by induction
on $k$. The induction base is trivial.  Assume
$\mbox{\rm S}(i,1)=\mbox{\rm S}(i+k-1,k)$ for $k>0$. We show that
\mbox{$\mbox{\rm S}(i,1)=\mbox{\rm S}(i+k,1+k)$}.  
For the first inclusion, assume
$L\in \mbox{\rm S}(i,1)$, and let $f$ be an $\mbox{\rm S}(i+k-1,k)$-selector
for $L$ that exists by the inductive hypothesis.  Given any distinct 
input strings $y_1,\ldots ,y_m$, $m\geq 1$, an
$\mbox{\rm S}(i+k,1+k)$-selector $g$ for $L$ is defined by
\[
g(y_1,\ldots ,y_m) \equalsdef  \left \{
\begin{array}{ll}
f(\{y_1,\ldots ,y_m\}-\{z\})\cup \{z\} & 
\mbox{if $f(y_1,\ldots ,y_m)\neq \emptyset$}\\
Y & \mbox{otherwise,}
\end{array}
\right.
\]
where $z\in f(y_1,\ldots ,y_m)$ and $Y$ is an arbitrary subset of
$\{y_1,\ldots ,y_m\}$. Clearly, $g\in \fp$, $g(y_1,\ldots ,y_m) \seq
\{y_1,\ldots ,y_m\}$, and if $\|L\cap \{y_1,\ldots ,y_m\}\| \geq i+k$,
then $g$ outputs at least $1+k$ strings each belonging to~$L$. Thus,
$L\in \mbox{\rm S}(i+k,1+k)$ via~$g$.

\smallskip

For the converse inclusion, let $L\in \mbox{\rm S}(i+k,1+k)$ via $g$.  To
define an \mbox{$\mbox{\rm S}(i+k-1,k)$}-selector $f$ for $L$, 
let $i+k$ strings
$z_1,\ldots ,z_{i+k}\in L$ (w.l.o.g., $L$ is infinite) be hard-coded
into the machine computing~$f$.
Given $y_1,\ldots ,y_m$ as input strings, $m\geq 1$, define 
$$ 
f(y_1,\ldots ,y_m) \equalsdef 
\left \{
\begin{array}{ll}
g(y_1,\ldots ,y_m) & 
\mbox{if $\{z_1,\ldots ,z_{i+k}\} \seq \{y_1,\ldots ,y_m\}$}\\
g(y_1,\ldots ,y_m,z) - \{z\} &
\mbox{otherwise,}
\end{array}
\right.
$$ 
where $z\in \{z_1,\ldots ,z_{i+k}\} - \{y_1,\ldots ,y_m\}$.  Clearly,
$f \in \fp$ selects a subset of its inputs \mbox{$\{y_1,\ldots
  ,y_m\}$}, and if $\|L\cap \{y_1,\ldots ,y_m\}\| \geq i+k-1$, then
$f$ outputs at least $k$ elements of~$L$. Thus, $f$ witnesses that
$L\in \mbox{\rm S}(i+k-1,k)$, which equals $\mbox{\rm S}(i,1)$ by the
inductive hypothesis.~\hspace*{\fill}$\Box$

\begin{proposition}
\label{fact:P-sel}
\begin{enumerate}
\item \label{fact-1}
$\mbox{\rm S}(1)=\psel$.

\item \label{fact-2}
$(\forall k\geq 1)\, [\mbox{\rm S}(k) \seq \mbox{\rm S}(k+1)]$.
\end{enumerate}
\end{proposition}

\noindent 
{\bf Proof.} \quad
By definition, we have immediately Part~\ref{fact-2} and the inclusion
from left to right in Part~\ref{fact-1}, as in particular, given any
pair of strings, an $\mbox{\rm S}(1)$-selector $f$ is required to select a
string (recall our assumption that all $\mbox{\rm S}(k)$-selectors output
{\em exactly\/} one input string) that is no less likely to be in the
set than the other one.  For the converse inclusion, fix any set of
inputs $y_1,\ldots ,y_m$, $m\geq 1$, and let $f$ be a P-selector for
$L$.  Play a knock-out tournament among the strings $y_1,\ldots ,y_m$,
where $x$ beats $y$ if and only if $f(x,y)=x$. %
Let $y_w$ be the winner. Clearly, $g(y_1,\ldots ,y_m) \equalsdef y_w$ 
witnesses that $L\in \mbox{\rm S}(1)$.  \hspace*{\fill}$\Box$

\medskip

Next we prove that SH is
properly infinite and is strictly contained in $\mbox{\rm Fair-S}(n-1,1)$.
Recall that, by convention, the ``$n-1$'' in
$\mbox{\rm Fair-S}(n-1,1)$\label{ind:fair-n-1} denotes the non-constant
threshold function \mbox{$g(n) = n - 1$}.  
Fix an enumeration $\{f_i\}_{i\geq 1}$ of FP functions, and define
$e(0) \equalsdef 2$ and $e(k) \equalsdef 2^{e(k-1)}$\label{ind:wide-e}
for $k \geq 1$. For each $i\geq 0$ and $s\leq 2^{e(i)}$, let $W_{i,s}
\equalsdef \{w_{i,1}, \ldots , w_{i,s}\}$\label{ind:wis} be an
enumeration of the lexicographically smallest $s$ strings in
$\Sigma^{e(i)}$ (this notation will be used also in
Section~\ref{sec:gch}).

\begin{theorem}
\label{thm:proper}
\begin{enumerate}
\item 
\label{proper:1}
For each $k\geq 1$, $\mbox{\rm S}(k) \subset \mbox{\rm S}(k+1)$.

\item 
\label{proper:2}
$\mbox{\rm SH} \subset \mbox{\rm Fair-S}(n-1,1)$.
\end{enumerate}
\end{theorem}

\noindent
{\bf Proof.} 
\ref{proper:1}. For fixed $k\geq 1$, choose $k+1$ pairwise distinct strings
$b_0,\ldots ,b_k$ of the same length.
Define
\[
A_k  \equalsdef  \bigcup_{i \geq 1} \left( 
\left\{ b_{0}^{e(i)},\ldots ,b_{k}^{e(i)} \right\} -
\left\{ f_{i}(b_{0}^{e(i)},\ldots ,b_{k}^{e(i)}) \right\}
\right),
\]
i.e., for each $i \geq 1$, $A_k$ can lack at most one out of the $k+1$
strings $b_{0}^{e(i)},\ldots ,b_{k}^{e(i)}$.

An $\mbox{\rm S}(k+1)$-selector $g$ for $A_k$ is given in 
Figure~\ref{figure:selector}.
W.l.o.g., assume each input in \mbox{$Y = \{ y_1,\ldots ,y_m \}$}
to be of the form
$b_{j}^{e(i)}$ for some $j\in \{0,\ldots , k\}$ and $i\in \{i_1,\ldots
,i_s\}$, where $1\leq i_1 < \cdots <i_s$ and $s\leq m$.
Clearly, $g(Y) \in Y$.  
Let $n = | \< y_1,\ldots ,y_m \> |$. 
Since there are at most $m$ while loops to be executed and
the polynomial-time transducers $f_{i_t}$, $t<s$, run on
inputs of length at most $c\cdot \log e(i_s)$ for some constant $c$,
the runtime of $g$ on {\em that\/} input is bounded above by some
polylogarithmic function in $n$.  Then, there is a polynomial in $n$
bounding $g$'s runtime on {\em any\/} input.  Thus, $g\in \fp$.  If some
element $y$ is output during the while loop, then $y\in A_k$.  If $g$
outputs an arbitrary input string after exiting the while loop, then
no input of the form $b_{j}^{e(i_t)}$, $t<s$, is in $A_k$, and since
$A_k$ has at most $k+1$ strings at each length, we have $\|A_k\cap Y\| 
\leq k$ if $g(Y) \not\in A_k$.
Thus, $A_k\in \mbox{\rm S}(k+1)$ via~$g$.

On the other hand, each potential $\mbox{\rm S}(k)$-selector $f_i$, given
$b_{0}^{e(i)},\ldots ,b_{k}^{e(i)}$ as input strings, outputs an
element not in $A_k$ though $k$ of these strings are in~$A_k$.  Thus,
\mbox{$A_k\not\in \mbox{\rm S}(k)$}.

\begin{figure}[tbp]
\begin{center}
{ \singlespacing 
\begin{construction}
\item {\bf Description of an} {\boldmath $\mbox{\bf S}(k+1)$}{\bf -selector} 
{\boldmath $g.$} %
  \begin{block}
    \item {\bf input} $Y=\{y_1,\ldots ,y_m\}$
    \item {\bf begin} $t:=s-1$;
    \begin{block}
       \item {\bf while} $t\geq 1$ {\bf do} 
       \begin{block}
          \item   $Z := \{ y\in Y \, | \, 
                  (\exists j\in \{0,\ldots ,k\})\, 
                  [y = b_{j}^{e(i_t)}] \}
                  - \{ f_{i_t}(b_{0}^{e(i_t)},\ldots
                  ,b_{k}^{e(i_t)}) \}$;
          \item {\bf if} $Z \neq \emptyset $ {\bf then output}
                some element of $Z$ and {\bf halt}
          \item {\bf else} $t:=t-1$
       \end{block}
       \item {\bf end while}
       \item {\bf output} an arbitrary input string and {\bf halt}
    \end{block}
    \item {\bf end}
  \end{block}
  \item {\bf End of description of {\boldmath $g$}.} 
\end{construction}

} %
\end{center}
\caption{\label{figure:selector} An $\mbox{\rm S}(k+1)$-selector $g$ for $A_k$.}
\end{figure}

\smallskip

\ref{proper:2}.  Fix any $k\geq 1$, and let $L\in
\mbox{\rm S}(k)$ via selector $f$. For each of the finitely many
tuples $y_1, \ldots , y_{\ell}$ such that $\ell \leq k$ and
$|y_i|\leq \ell$, $1\leq i\leq \ell$, let $z_{y_1, \ldots , y_{\ell}}$
be some fixed string in $L\cap \{y_1, \ldots , y_{\ell} \}$ if this set is
non-empty, and an arbitrary string from $\{y_1, \ldots , y_{\ell} \}$
otherwise. Let these fixed strings be hard-coded into the machine
computing the function $g$ defined by
$$ 
g(y_1,\ldots ,y_n) \equalsdef 
\left \{
\begin{array}{ll}
\{z_{y_1, \ldots , y_n}\} & \mbox{if $n\leq k$}\\
\{f(y_1,\ldots ,y_n)\}    & \mbox{otherwise.}
\end{array}
\right. 
$$ 
Thus, $L\in \mbox{\rm Fair-S}(n-1,1)$ via $g$, showing that $\mbox{\rm SH}
\seq \mbox{\rm Fair-S}(n-1,1)$.  

The strictness of the inclusion is proven
as in Part~\ref{proper:1} of this proof. To define a set $A \not\in
\mbox{\rm SH}$ we have here to diagonalize against all potential selectors
$f_j$ and all levels of SH simultaneously.  That is, in stage $i = \<
j,k \>$ of the construction of $A  \equalsdef  \bigcup_{i \geq 1} A_i$, we will
diagonalize against $f_j$ being an $\mbox{\rm S}(k)$-selector for $A$. Fix
$i = \< j,k \>$. Recall that $W_{i,k+1}$ is the set of the smallest
$k+1$ length $e(i)$ strings. Note that $2^{e(i)} \geq k+1$ holds for
each $i$, since we can w.l.o.g.\ assume that the pairing function
satisfies $u > \max \{v,w\}$ for all $u,v$, and $w$ with $u = \< v,w
\>$. Define $A_i  \equalsdef  W_{i,k+1} - \{f_j(W_{i,k+1})\}$.  Assume $A \in
\mbox{\rm SH}$, i.e., there exists some $t$ such that $A \in \mbox{\rm S}(t)$
via some selector $f_s$. But this contradicts that for $r = \< s,t
\>$, by construction of $A$, we have $\|A \cap W_{r,t+1}\| \geq t$,
yet $f_s(W_{r,t+1})$ either doesn't output one of its inputs (and is
thus no selector), or $f_s(W_{r,t+1}) \not\in A$. Thus, $A \not\in
\mbox{\rm SH}$.  

Now we prove that $A$ trivially is in
$\mbox{\rm Fair-S}(n-1,1)$, as $A$ is constructed such that the promise is
{\em never\/} met.  By way of contradiction, suppose a set $X$ of
inputs is given, \mbox{$\|X\| = n$}, \mbox{$\|A \cap X\| \geq n-1$}, 
and $|x| \leq n$ for each $x \in X$. 
Let $e(i)$ be the maximum length of the strings
in $A \cap X$, i.e., \mbox{$A \cap X = \bigcup_{m=1}^{i} A_m \cap X$}.  Let
$j$ and $k$ be such that $i = \< j,k \>$.  Since (by the above remark
about our pairing function) $k+1 \leq i$, we have by construction of
$A$,
\[
e(i) - 1 \leq n - 1 \leq \|A \cap X\| = \|\bigcup_{m=1}^{i} A_m  \cap X\| 
\leq \|\bigcup_{m=1}^{i} A_m\| \leq (k+1)i \leq i^2,
\]
which is false for all $i \geq 0$. Hence, $A \in \mbox{\rm Fair-S}(n-1,1)$.
\hspace*{\fill}$\Box$

\medskip

A variation of this technique proves that, unlike \psel, none of the
$\mbox{\rm S}(k)$ for $k\geq 2$ is closed under $\leq^p_m$-reductions.  (Of
course, every class $\mbox{\rm S}(k)$ is closed downwards under
polynomial-time one-one reductions.)\  We also show that sets in $\mbox{\rm S}(2)$
that are many-one reducible to their complements may already go beyond~P, 
which contrasts with Selman's result that a set $A$ is in P if
and only if \mbox{$A \leq_{m}^{p} \overline{A}$} and $A$ is 
P-selective~\cite{sel:j:pselective-tally}.  
It follows that the class P cannot
be characterized by the auto-reducible sets 
(see~\cite{buh-tor:j:pselective}) in any of the higher
classes in SH\@. It would be interesting to strengthen
Corollary~\ref{cor:auto-red} to the case of the {\em self\/}-reducible
sets, as that would contrast sharply with Buhrman and
Torenvliet's characterization of P as those self-reducible sets that
are in P-Sel~\cite{buh-tor:j:pselective}.

\begin{theorem}
\label{thm:separation}
\begin{enumerate}
\item
\label{separation}
For each $k\geq 2$, $\mbox{\rm S}(k) \subset {\Re}^p_m(\mbox{\rm S}(k))$.

\item
\label{auto-red}
There exists a set $A$ in $\mbox{\rm S}(2)$ such that $A \leq_{m}^{p}
\overline{A}$ and yet $A\not\in \p$.
\end{enumerate}
\end{theorem}

\begin{corollary}
\label{cor:auto-red} \quad
There exists an auto-reducible set in $\mbox{\rm S}(2)$ that is not in \p.
\end{corollary}

\noindent 
{\bf Proof of Theorem~\ref{thm:separation}.} 
\ref{separation}. In fact, for fixed~$k$, we will define a set 
$L$ in \mbox{${\Re}^p_m(\mbox{\rm S}(2))-\mbox{\rm S}(k)$}. 
By Fact~\ref{fact:P-sel}, the theorem follows.  
Choose $2k$ pairwise distinct strings $b_1, \ldots, b_{ 2k}$ 
of the same length. Define 
$L \equalsdef \bigcup_{i \geq 1} A_i \cup B_i$, where
\[
A_i \equalsdef \left\{
\begin{array}{ll}
\{ b_{1}^{e(i)},\ldots ,b_{k}^{e(i)} \} & 
\mbox{if $f_i(b_{1}^{e(i)},\ldots ,b_{2k}^{e(i)})\not\in
     \{ b_{1}^{e(i)},\ldots ,b_{k}^{e(i)} \} $} \\
\emptyset & \mbox{otherwise,}
\end{array}
\right.
\]
\[
B_i \equalsdef \left\{
\begin{array}{ll}
\{ b_{k+1}^{e(i)},\ldots ,b_{2k}^{e(i)} \} & 
\mbox{if $f_i(b_{1}^{e(i)},\ldots ,b_{2k}^{e(i)})\not\in
     \{ b_{k+1}^{e(i)},\ldots ,b_{2k}^{e(i)} \} $} \\
\emptyset & \mbox{otherwise.}
\end{array}
\right.
\]
Clearly, each potential $\mbox{\rm S}(k)$-selector $f_i$, given
$b_{1}^{e(i)},\ldots ,b_{2k}^{e(i)}$ as input strings, outputs an
element not in $L$ though $\| L\cap \{ b_{1}^{e(i)},\ldots
,b_{2k}^{e(i)} \}\| \geq k$.  Thus, $L\not\in \mbox{\rm S}(k)$.  

Now define the set 
\[
L'  \equalsdef  \{b^{e(i)}_1\,|\, b^{e(i)}_1\in L\}\cup \{b^{e(i)}_{k+1}\,|\,
b^{e(i)}_{k+1}\in L\}
\] 
and an FP function $g$ by \mbox{$g(b^{e(i)}_j)  \equalsdef 
b^{e(i)}_1$} if \mbox{$1\le j\le k$}, and 
\mbox{$g(b^{e(i)}_j)  \equalsdef  b^{e(i)}_{k+1}$} if
\mbox{$k+1\le j\le 2k$}, and \mbox{$g(x) = x$} for all $x$ not of the form
$b^{e(i)}_j$ for any \mbox{$i\geq 1$} and~$j$, $1\leq j\leq 2k$. Then, we
have $x\in L$ if and only if $g(x)\in L'$ for each $x \in
\sigmastar$, that is, $L\le^p_m L'$.  

Now we show that $L' \in \mbox{\rm S}(2)$.  Given any distinct inputs
$y_1,\ldots ,y_n$ (each having, without loss of generality, the form
$b^{e(i)}_1$ or $b^{e(i)}_{k+1}$ for some $i\geq 1$), define an
$\mbox{\rm S}(2)$-selector as follows:
\begin{description}
\item[Case 1:] All inputs have the same length. Then, $\{
  y_1,\ldots ,y_n \} \seq \{ b^{e(i)}_1, b^{e(i)}_{k+1}
  \}$ for some $i\geq 1$. Define $f(y_1,\ldots ,y_n)$ to
  be $b^{e(i)}_1$ if $b^{e(i)}_1\in \{y_1,\ldots,y_n\}$,
  and to be $b^{e(i)}_{k+1}$ otherwise.  Hence, $f$
  selects a string in $L'$ if $\|\{y_1,\ldots,y_n\}\cap
  L'\|\ge 2$.

\item[Case 2:] The input strings have different lengths. Let
  $\ell  \equalsdef  \max\{ |y_1|, \ldots , |y_n| \}$.  By brute
  force, we can decide in time polynomial in $\ell$ if
  there is some string with length smaller than $\ell$
  in $L'$. If so, $f$ selects the first string found.
  Otherwise, by the argument of Case~1, we can show that
  $f$ selects a string (of maximum length) in $L'$ if
  $L'$ contains two of the inputs.
\end{description}

\smallskip

\ref{auto-red}. Let $\{M_i\}_{i\geq 1}$ be an enumeration of all
deterministic polynomial-time Turing machines. Define
\[
A  \equalsdef  \{ 0^{e(i)} \,|\, i\geq 1\, \wedge \, 0^{e(i)}\not\in L(M_i)\}
\cup \{ 1^{e(i)} \,|\, i\geq 1\, \wedge \, 0^{e(i)}\in L(M_i)\}.
\]
Assume $A\in \p$ via $M_j$ for some $j\geq 1$. This contradicts that
$0^{e(j)}\in A$ if and only if $0^{e(j)}\not\in L(M_j)$. Hence,
$A\not\in \p$. Define an \fp\ function $g$ by
$g(0^{e(i)})  \equalsdef  1^{e(i)}$ and $g(1^{e(i)})  \equalsdef  0^{e(i)}$ 
for each $i\geq 1$; and for each $x\not\in \{0^{e(i)},1^{e(i)}\}$, define
$g(x)  \equalsdef  y$, where $y$ is a fixed string in $A$ (w.l.o.g.,
$A\neq \emptyset$). Clearly, $A \leq_{m}^{p} \overline{A}$ via $g$.
$A\in \mbox{\rm S}(2)$ follows as above.~\hfill$\Box$

\begin{definition}
\label{def:many-one-lexinc-red} \quad
For sets $A$ and $B$, $A \leq_{m,\, \ell i}^{p} B$ if there is an \fp\
function $f$ such that for all $x \in \sigmastar$, (a)~$x \in A \Lolra f(x)
\in B$, and (b)~$x <_{\mbox{\protect\scriptsize lex}} f(x)$.
\end{definition}

Note that a similar kind of reduction was defined and was of use in
\cite{hem-hoe-sie-you:j:enum}, and that,
intuitively, sets in $\{ L \,|\, L \leq_{m,\, \ell i}^{p} L \}$ may be
viewed as having a very weak type of padding functions.

\begin{theorem}
\label{thm:paddable} \quad
If $L \in \mbox{\rm SH}$ and $L \leq_{m,\, \ell i}^{p} L$, then
$L \in \psel$.
\end{theorem}

\noindent
{\bf Proof.} 
Let $L \leq_{m,\, \ell i}^{p} L$ via $f$, and let $g$ be an
$\mbox{\rm S}(k)$-selector for $L$, for some $k$ for which $L \in
\mbox{\rm S}(k)$.  A P-selector $h$ for $L$ is defined as follows: Given
any inputs $x$ and $y$, generate two chains of $k$ lexicographically
increasing strings by running the reduction~$f$, i.e., 
\mbox{$x = x_1 <_{\mbox{\protect\scriptsize lex}} x_2 
<_{\mbox{\protect\scriptsize lex}} \cdots
<_{\mbox{\protect\scriptsize lex}} x_k$} and 
\mbox{$y = y_1 <_{\mbox{\protect\scriptsize lex}} y_2 
<_{\mbox{\protect\scriptsize lex}} \cdots 
<_{\mbox{\protect\scriptsize lex}} y_k$}, where 
$x_2 = f(x)$, $x_3 = f(f(x))$, etc., and similarly for the~$y_i$.  
To ensure that $g$ will run on {\em distinct\/} inputs only
(otherwise, $g$ is not obliged to meet requirements 1 and 2 of
Definition~\ref{def:S-fair-S}), let $z_1, \ldots , z_l$ be all the
$y_i$'s not in $\{x_1, \ldots , x_k\}$.  Now run $g(x_1 , \ldots ,
x_k, z_1, \ldots , z_l)$ and define $h(x,y)$ to output $x$ if $g$
outputs some string $x_i$, and to output $y$ if $g$ selects some
string $y_i$ (recall our assumption that $\mbox{\rm S}(k)$-selectors such
as $g$ output exactly one string). Clearly, $h \in \fp$, and if $x$ or
$y$ are in $L$, then at least $k$ inputs to $g$ are in $L$, so $h$
selects a string in $L$.~\hfill$\Box$

\medskip

Theorem~\ref{thm:proper} and Theorem~\ref{thm:paddable} immediately
imply the following:

\begin{corollary}
\label{cor:paddable} 
\quad
$\mbox{\rm SH} \not\seq \{ L \,|\, L \leq_{m,\, \ell i}^{p} L \}$.
\end{corollary}

Ogihara \cite{ogi:j:comparable} has recently introduced the
polynomial-time membership-comparable sets 
as another generalization of the P-selective sets.

\begin{definition}
\label{def:p-mc}
\cite{ogi:j:comparable} 
\quad 
Let $g$ be a monotone non-decreasing and polynomially bounded FP
function from $\N$ to~$\N^{+}$.
\begin{enumerate}
\item A function $f$ is called a $g$-membership-comparing function (a
  $g$-mc-function, for short) for $A$ if for every $z_1,\ldots ,z_m$
  with $m \geq g(\max\{|z_1|,\ldots ,|z_m|\})$,
\begin{eqnarray*}
f(z_1,\ldots ,z_m)\in \{0,1\}^m & \mbox{and} & 
(\chi_A(z_1),\ldots ,\chi_A(z_m)) \neq f(z_1,\ldots ,z_m).
\end{eqnarray*}

\item A set $A$ is polynomial-time $g$-membership-comparable if there
  exists a polynomial-time computable $g$-mc-function for $A$.

\item P-mc($g$) denotes the class of polynomial-time
  $g$-membership-comparable sets.

\item $\mbox{\rm P-mc}(\mbox{const}) \equalsdef \bigcup\{\mbox{\rm P-mc}(k)\,|\,
  k\geq 1\}$, $\mbox{\rm P-mc}(\mbox{log}) \equalsdef
  \bigcup\{\mbox{\rm P-mc}(f)\,|\, f\in {\cal O}(\log)\}$, and
  $\mbox{\rm P-mc}(\mbox{poly}) \equalsdef \bigcup\{\mbox{\rm P-mc}(p)\,|\, p
  \mbox{ is a polynomial} \}$.
\end{enumerate}
\end{definition}

\begin{remark} \quad
  We can equivalently (i.e., without changing the class) require in
  the definition that $f(z_1,\ldots ,z_m) \neq (\chi_A(z_1),\ldots
  ,\chi_A(z_m))$ must hold only if the inputs \mbox{$z_1,\ldots ,z_m$}
  happen to be {\em distinct}.  This is true because if there are $r$
  and $t$ with $r \neq t$ and $z_r = z_t$, then $f$ simply outputs a
  length $m$ string having a ``0'' at position $r$ and a ``1'' at
  position $t$.
\end{remark}

\begin{figure}[tbp]
\centerline{\psfig{figure=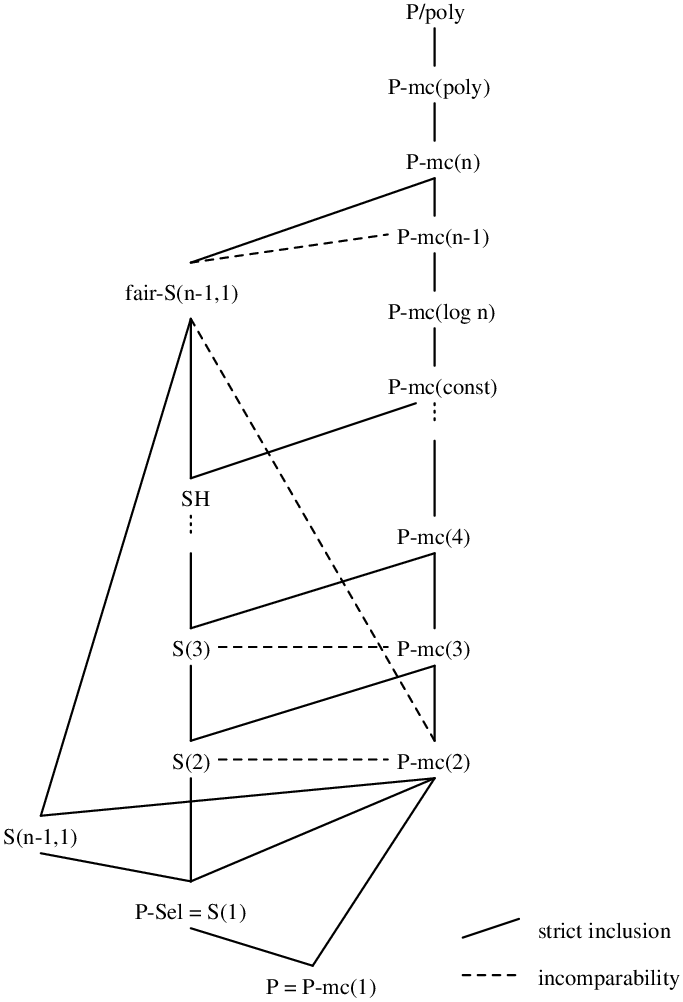,width=4.8in,height=6in}}%
\caption{Inclusion relationships among S, Fair-S, and P-mc classes.\label{f:inclusion-relations}}%
\end{figure}

Since $\mbox{\rm P-mc}(k)$ is closed under
$\leq_{1\mbox{-}tt}^{p}$-reductions for each~$k$~\cite{ogi:j:comparable} 
but none of the $\mbox{\rm S}(k)$ for $k\geq 2$ is
closed under $\leq^p_m$-reductions (Theorem~\ref{thm:separation}), it
is clear that Ogihara's approach to generalized selectivity is
different from ours, and in Theorem~\ref{thm:fair-s-p-mc} below, we
completely establish, in terms of incomparability and strict
inclusion, the relations between his and our generalized selectivity
classes (see Figure~\ref{f:inclusion-relations}).
Note that Part~\ref{thm:fair-s-p-mc:2} of Theorem~\ref{thm:fair-s-p-mc} 
generalizes to $k$ larger than 1 a result of Ogihara---who
proved that the P-selective sets are strictly contained in
$\mbox{\rm P-mc}(2)$~\cite{ogi:j:comparable}---and the known fact
that P-Sel is strictly larger than~P~\cite{sel:j:pselective-tally}.

\begin{theorem}
\label{thm:fair-s-p-mc}
\begin{enumerate}
\item\label{thm:fair-s-p-mc:1}
$\mbox{\rm P-mc}(2)\not\seq \mbox{\rm Fair-S}(n-1,1)$.

\item\label{thm:fair-s-p-mc:2}
For each $k\geq 1$, $\mbox{\rm S}(k) \subset \mbox{\rm P-mc}(k+1)$ and
$\mbox{\rm S}(k) \not\seq \mbox{\rm P-mc}(k)$\@.

\item\label{thm:fair-s-p-mc:3}
$\mbox{\rm S}(n-1,1) \subset \mbox{\rm P-mc}(2)$.

\item\label{thm:fair-s-p-mc:4}
$\mbox{\rm Fair-S}(n-1,1) \subset \mbox{\rm P-mc}(n)$ and
$\mbox{\rm Fair-S}(n-1,1) \not\seq \mbox{\rm P-mc}(n-1)$.
\end{enumerate}
\end{theorem}

\noindent 
{\bf Proof.} \quad
First recall that $\{f_i\}_{i\geq 1}$ is our enumeration
of FP functions and that the set 
\mbox{$W_{i,s} = \{w_{i,1}, \ldots , w_{i,s}\}$},
for $i\geq 0$ and $s \leq 2^{e(i)}$,
collects the lexicographically smallest $s$ 
strings in~$\Sigma^{e(i)}$, where function $e$ is inductively defined
to be $e(0) = 2$ and $e(i) = 2^{e(i-1)}$ for $i\geq 1$.  Recall also
our assumption that a selector for a set in SH outputs a single input
string (if the promise is met), whereas $\mbox{\rm S}(n-1,1)$ and
$\mbox{\rm Fair-S}(n-1,1)$ are defined via selectors that may output subsets
of the given set of inputs.

\smallskip

\ref{thm:fair-s-p-mc:1}.  We will construct a set $A$ in stages. Let
$u_i$ be the smallest string in \mbox{$W_{i,e(i)} \cap f_i(W_{i,e(i)})$} (if
this set is non-empty; otherwise, $f_i$ immediately disqualifies for
being a $\mbox{\rm Fair-S}(n-1,1)$-selector and we may go to the next
stage).  Define 
\[
A \equalsdef \bigcup_{i\geq 1} (W_{i,e(i)} - \{u_i\}).
\] 
Then, $A\not\in \mbox{\rm Fair-S}(n-1,1)$, since for any~$i$,
$f_i(W_{i,e(i)})$ outputs a string not in $A$ although $e(i) - 1$ of
these inputs (each of length~$e(i)$, i.e., the inputs satisfy the
``fairness condition'') {\em are\/} in~$A$.

For defining a P-mc(2) function $g$ for $A$, let any distinct inputs
$y_1,\ldots , y_m$ with $m\geq 2$ be given.  If there is some $y_j$
such that $y_j \not\in W_{i,e(i)}$ for each~$i$, then define
$g(y_1,\ldots , y_m)$ to be $0^{j-1}10^{m-j}$. If there is some $y_j$ with
$|y_j| < e(i_0)$, where $e(i_0) = \max\{|y_1|,\ldots , |y_m|\}$, then
compute the bit $\chi_{\overline{A}}(y_j)$ by brute force in time
polynomial in $e(i_0)$, and define $g(y_1, \ldots , y_m)$ to be
$0^{j-1} \chi_{\overline{A}}(y_j) 0^{m-j}$.  Otherwise (i.e., if
$\{y_1,\ldots , y_m\} \seq W_{i_0,e(i_0)}$), let $g(y_1,\ldots , y_m)$
be~$0^m$.  Since, by definition of~$A$, there is at most one string in
$W_{i_0,e(i_0)}$ that is not in~$A$, but $m\geq 2$, we have
$g(y_1,\ldots , y_m) \neq (\chi_A(y_1),\ldots , \chi_A(y_m))$.  Thus,
$A\in \mbox{\rm P-mc}(2)$ via~$g$.

\smallskip

\ref{thm:fair-s-p-mc:2}. For fixed $k\geq 1$, let $L\in \mbox{\rm S}(k)$
via~$f$. Define a $\mbox{\rm P-mc}(k+1)$ function $g$ for $L$ that, given
distinct inputs $y_1,\ldots ,y_m$ with $m\geq k+1$, outputs the string
$1^{j-1}01^{m-j}$ if $y_j$ is the string output by $f(y_1,\ldots
,y_m)$. Clearly, $g(y_1,\ldots ,y_m) \neq (\chi_L(y_1),\ldots
,\chi_L(y_m))$, since there are at least $k$ 1's in $1^{j-1}01^{m-j}$,
and $f(y_1,\ldots ,y_m) = y_j$ is thus a string in~$L$.  Hence, $L\in
\mbox{\rm P-mc}(k+1)$ via~$g$, showing $\mbox{\rm S}(k) \seq \mbox{\rm P-mc}(k+1)$.
By Statement~\ref{thm:fair-s-p-mc:1}, this inclusion is strict, and so
is any inclusion to be proven below.  

To show that $\mbox{\rm S}(k) \not\seq \mbox{\rm P-mc}(k)$, fix $k$ strings
$b_1,\ldots ,b_k$ of the same length. Define 
\[ 
A  \equalsdef  \left\{
b_{j}^{e(i)}\,
\begin{tabular*}{6.8cm}{|l}
\ $i\geq 1$ and
$f_i(b_{1}^{e(i)},\ldots ,b_{k}^{e(i)}) \in \{0,1\}^k$\\
\ and has a ``1'' at position $j,\ 1\leq j\leq k $ 
\end{tabular*}
\right\} .
\]
Clearly, since $f_i(b_{1}^{e(i)},\ldots ,b_{k}^{e(i)}) =
(\chi_A(b_{1}^{e(i)}),\ldots ,\chi_A(b_{k}^{e(i)}))$ for each~$i$, no
\fp\ function $f_i$ can serve as a $\mbox{\rm P-mc}(k)$ function for $A$.
To define an $\mbox{\rm S}(k)$-selector for $A$, let any inputs $y_1,
\ldots , y_m$ (w.l.o.g., each of the form $b_{j}^{e(i)}$) be given,
and let $\ell = \max\{|y_1|, \ldots , |y_m|\}$. As in the proofs of
Theorem~\ref{thm:proper} and Theorem~\ref{thm:separation}, it can be
decided in time polynomial in $\ell$ whether there is some string of
length smaller than $\ell$ in $A$. If so, the $\mbox{\rm S}(k)$-selector
$f$ for $A$ selects the first such string found. Otherwise, $f$
outputs an arbitrary string of maximum length.  Since there are at
most $k$ strings in $A$ at any length, either the output string is in
$A$, or $\| A \cap \{y_1, \ldots , y_m\}\| < k$. Thus, $\mbox{\rm S}(k)
\not\seq \mbox{\rm P-mc}(k)$.  Statement~\ref{thm:fair-s-p-mc:1} implies
that as well $\mbox{\rm P-mc}(k)\not\seq \mbox{\rm S}(k)$ for $k\geq 2$; the
$k$th level of $\mbox{\rm SH} = \bigcup_{i\geq 1} \mbox{\rm S}(i)$ and the
$k$th level of the
hierarchy within $\mbox{\rm P-mc}(\mbox{const})$ are thus incomparable.

\smallskip

\ref{thm:fair-s-p-mc:3}. Let $L\in \mbox{\rm S}(n-1,1)$ via selector $f$.
Define a P-mc(2) function $g$ for $L$ as follows: Given distinct input
strings $y_1,\ldots ,y_n$ with $n\geq 2$, $g$ simulates $f(y_1,\ldots
,y_n)$ and outputs the string $1^{j-1}01^{n-j}$ if $y_j$ is any (say
the smallest) string in $f(y_1,\ldots ,y_n)$. Again, we can exclude
one possibility for $(\chi_A(y_1),\ldots ,\chi_A(y_n))$ via $g$ in
polynomial time, because the $\mbox{\rm S}(n-1,1)$-promise is met for the
string $1^{j-1}01^{n-j}$, and thus $f$ must output a string in $L$.

\smallskip

\ref{thm:fair-s-p-mc:4}. Now we show that the proof of
Statement~\ref{thm:fair-s-p-mc:3} fails to some extent for the
corresponding Fair-class, i.e., we will show that
$\mbox{\rm Fair-S}(n-1,1) \not\seq \mbox{\rm P-mc}(n-1)$.  This resembles
Part~\ref{thm:fair-s-p-mc:2} of this theorem, but note that the proof
now rests also on the ``fairness condition'' rather than merely on
the~$(n-1)$-promise.  We also show that the ``fairness condition'' can
no longer ``protect'' $\mbox{\rm Fair-S}(n-1,1)$ from being contained
in~$\mbox{\rm P-mc}(n)$.

$A \equalsdef \bigcup_{i\geq 1} A_i$ is defined in stages so that in
stage~$i$, $f_i$ fails to be a $\mbox{\rm P-mc}(n-1)$ function for~$A_i$.
This is ensured by defining $A_i$ as a subset of the $e(i)-1$ smallest
strings of length~$e(i)$, $W_{i,e(i)-1}$, such that $w_{i,j}\in A_i$
if and only if $f_i(W_{i,e(i)-1})$ outputs a string of length $e(i)-1$
and has a ``1'' at position~$j$. Thus, $A$ is not in $\mbox{\rm P-mc}(n-1)$,
since \mbox{$f_i(w_{i,1}, \ldots , w_{i,e(i)-1}) =
  (\chi_{A}(w_{i,1}),\ldots ,\chi_{A}(w_{i,e(i)-1}))$} for each $i\geq
1$.

To see that $A\in \mbox{\rm Fair-S}(n-1,1)$, let any distinct inputs
$y_1,\ldots , y_n$ be given, each having, w.l.o.g., length $e(i)$ for
some $i$, and let $e(i_0)$ be their maximum length.  As before, if
there exists a string of length smaller than $e(i_0)$, say $y_j$, then
it can be decided by brute force in polynomial time whether or not
$y_j$ belongs to $A$.  Define a $\mbox{\rm Fair-S}(n-1,1)$-selector~$g$ to
output $\{y_j\}$ if $y_j\in A$, and to output any input different from
$y_j$ if $y_j \not\in A$.  Thus, either the string output by $g$ does
belong to $A$, or \mbox{$\|A\cap \{ y_1,\ldots , y_n \}\| < n-1$}. On the
other hand, if all input strings are of the same length $e(i_0)$ and
\mbox{$\{y_1,\ldots , y_n\} \seq W_{i_0,e(i_0)-1}$}, then the ``fairness
condition'' is not fulfilled, as $e(i_0) > n$, and $g$ is thus not
obliged to output a string in $A$. If all inputs have length $e(i_0)$
and \mbox{$\{y_1,\ldots , y_n\} \not\seq W_{i_0,e(i_0)-1}$}, then by the
above argument, $g$ can be defined such that either the string output
by $g$ does belong to $A$, or $\|A\cap \{ y_1,\ldots , y_n \}\| <
n-1$. This completes the proof of $A\in \mbox{\rm Fair-S}(n-1,1)$.

Finally, we show that $\mbox{\rm Fair-S}(n-1,1) \seq \mbox{\rm P-mc}(n)$. Let
$L$ be a set in $\mbox{\rm Fair-S}(n-1,1)$ via selector~$f$. 
Let $y_1,\ldots , y_n$
be any distinct input strings such that \mbox{$n \geq \max \{|y_1|,\ldots ,
|y_n|\}$}, i.e., the ``fairness condition'' is now satisfied. Define a
P-mc-function $g$ for $L$ which, on inputs $y_1,\ldots , y_n$,
simulates $f(y_1,\ldots , y_n)$ and outputs the string
$1^{j-1}01^{n-j}$ if $f$ selects~$y_j$. Thus,
$$g(y_1,\ldots , y_n) \neq (\chi_L(y_1),\ldots ,\chi_L(y_n)),$$ 
and we have $L\in \mbox{\rm P-mc}(n)$ via~$g$.~\hfill$\Box$

\subsection{Circuit, Lowness, and Collapse Results}
\label{subsec:circuit}

This section demonstrates that the core results (i.e., small circuit,
$\mbox{\rm Low}_2$-ness, and collapse results) that hold for the P-selective
sets and that prove them structurally simple also hold for our
generalized selectivity classes.

Since $\mbox{\rm P-mc}(\mbox{poly}) \seq \ppoly$ \cite{ogi:j:comparable}
and $\mbox{\rm Fair-S}(n-1,1)$ is by Theorem~\ref{thm:fair-s-p-mc}
(strictly) contained in $\mbox{\rm P-mc}(n)$, it follows immediately that
every set in \mbox{$\mbox{\rm Fair-S}(n-1,1)$} has polynomial-size circuits and
is thus in EL$\Theta_3$ (by K\"obler's result that \mbox{$\ppoly \seq
\mbox{\rm EL}{\Theta}_3$}~\cite{koe:j:locating}).  Note that Ogihara refers
to Amir, Beigel, and Gasarch, whose \ppoly\ proof for
``non-p-superterse'' sets
(see~\cite[Theorem~10]{ami-bei-gas:conf-plus-manu:uni}) applies to Ogihara's
class $\mbox{\rm P-mc}(\mbox{poly})$ as well.  On the other hand,
\mbox{P-selective} \np\ sets can even be shown to be in
$\mbox{\rm Low}_2$~\cite{ko-sch:j:circuit-low}, the second level of the
low hierarchy within~NP\@. In contrast, the proof
of~\cite[Theorem~10]{ami-bei-gas:conf-plus-manu:uni} does not give a
$\mbox{\rm Low}_2$-ness result for non-p-superterse \np\ sets, and thus
also does not provide such a result for $\mbox{\rm P-mc}(\mbox{poly})\cap
\np$.  By modifying the technique of Ko and Sch\"oning, however, we
generalize in Theorem~\ref{thm:lowness} their result to our larger
selectivity classes.
Very recently, 
K\"obler~\cite{koe:c:low-survey} has observed that
our generalization of Ko and Sch\"oning's result that
\mbox{$\psel \cap \np \seq \mbox{\rm Low}_2$} can be combined with others 
to yield a very generalized statement. In particular, he observed that
our technique for proving Theorem~\ref{thm:lowness} and 
the techniques used to prove results such as 
``any P-cheatable NP set is 
$\mbox{\rm Low}_2$''~\cite{ami-bei-gas:conf-plus-manu:uni}
and 
``any NPSV-selective NP set is
$\mbox{\rm Low}_2$''~\cite{hem-nai-ogi-sel:j:refinements}
are compatible.  By combining the generalizing techniques
simultaneously, K\"obler can claim:
Any NP set
that is ``strongly membership-comparable by NPSV functions'' is
$\mbox{\rm Low}_2$~\cite{koe:c:low-survey}. 
(For the notations not defined here, we refer
to~\cite{koe:c:low-survey,ami-bei-gas:conf-plus-manu:uni,hem-nai-ogi-sel:j:refinements}.)

The proof of Theorem~\ref{thm:lowness} explicitly
constructs a family of non-uniform advice sets for any set in
$\mbox{\rm Fair-S}(n-1,1)$, as merely stating the existence of those
advice sets (which follows from Theorem~\ref{thm:small-circuits}) does
not suffice for proving $\mbox{\rm Low}_{2}$-ness.

Note that some results of this section (e.g.,
Theorem~\ref{thm:small-circuits}) extend to the more general GC
classes that will be defined in Section~\ref{sec:gch}. We propose as
an interesting task to explore whether all results of this section,
in particular the $\mbox{\rm Low}_2$-ness result of Theorem~\ref{thm:lowness},
apply to the GC classes.

\begin{theorem}
\label{thm:small-circuits} \quad
$\mbox{\rm Fair-S}(n-1,1) \seq \ppoly$.
\end{theorem}

\begin{corollary}\label{cor:small-circuits} \quad
$\mbox{\rm SH} \seq \ppoly$.
\end{corollary}

\begin{corollary}\label{cor2:small-circuits} \quad
$\mbox{\rm Fair-S}(n-1,1) \seq \mbox{\rm EL}{\Theta}_3$.
\end{corollary}

\begin{theorem}\label{thm:lowness} \quad
Any set in $\np \cap \mbox{\rm Fair-S}(n-1,1)$ is $\mbox{\rm Low}_2$.
\end{theorem}

\noindent
{\bf Proof.} \quad
Let $L$ be any \np\ set in $\mbox{\rm Fair-S}(n-1,1)$, and
let $f$ be a selector for $L$ and $N$ be an NPM such that $L=L(N)$.
First, for each length $m$, we shall construct a polynomially
length-bounded advice $A_m$ that helps deciding membership of any
string~$x$, $|x|=m$, in $L$ in polynomial time. For $m < 4$, take $A_m
 \equalsdef  L^{=m}$ as advice. From now on let $m\geq 4$ be fixed, 
and let $n$ be such that $4  \leq 2n \leq m$.  

Some notations are in order. A subset $G$ of $L^{=m}$ is called a {\em
  game\/} if $\|G\| = n$. Any output $w\in f(G)$ is called a {\em
  winner\/} of game $G$, and is said to be {\em yielded\/} by the {\em
  team\/} $G - \{w\}$.  If \mbox{$\|L^{=m}\| \leq 2(n+1)$}, then
simply take $A_m \equalsdef L^{=m}$ as advice. Otherwise, $A_m$ is
constructed in {\em rounds}.  In round~$i$, one team, $t_i$, is added
to $A_m$, and all winners yielded by that team in any game are deleted
from a set $B_{i-1}$. Initially, $B_0$ is set to be~$L^{=m}$.

In more detail, in the first round, all games of \mbox{$B_0 =
  L^{=m}$}, one after the other, are fed into the selector~$f$ for~$L$
to determine all winners of each game, and, associated with each
winner, the team yielding that winner. We will argue below that there
must exist at least one team yielding at least
$\frac{{N \choose n}}{{N \choose n-1}}$ winners if~$N$ is the number of
strings in~$L^{=m}$. Choose the ``smallest'' (according to the
ordering $\leq_{\mbox{\protect\scriptsize lex}}$ 
on~$L^{=m}$) such team, $t_1$, and add it to the advice~$A_m$.  Delete
from~$B_0$ all winners yielded by~$t_1$ and set~$B_1$ to be the
remainder of~$B_0$, i.e.,
\[
B_1 \equalsdef B_0 - \{ w \,|\,
  \mbox{winner $w$ is yielded by team $t_1$} \},
\]
and, entering the second round, repeat this procedure with all games
of~$B_1$ unless \mbox{$B_1$} has $\leq 2(n+1)$ elements. 
In the second round, a second
team~$t_2$, and in later rounds more teams~$t_i$, are determined and
are added to~$A_m$. The construction of~$A_m$ in rounds will terminate
if $\|B_{k(m)}\| \leq 2(n+1)$ for some integer $k(m)$ depending on the
given length~$m$. In that case, add $B_{k(m)}$ to~$A_m$.  Formally,
$$
A_m  \equalsdef B_{k(m)} \cup \bigcup_{i=1}^{k(m)} t_i,
$$
where $B_{k(m)} \seq L^{=m}$ contains at most \mbox{$2(n+1)$} elements,
\mbox{$t_i \seq L^{=m}$} is the team added to~$A_m$ in round~$i$, $1
\leq i \leq k(m)$, and the bound $k(m)$ on the number of rounds
executed at length $m$ is specified below.

We now show that there is some polynomial in~$m$ 
bounding the length of (the coding of) $A_m$ for any~$m$.  If
$L^{=m}$ has $N > 2(n+1)$ strings, then there are ${N \choose n}$
games and ${N \choose n-1}$ teams in the first round.  Since every
game has at least one winner, there exists one team yielding at least
$$
\frac{{N \choose n}}{{N \choose n-1}} 
= \frac{N-n+1}{n} > \frac{N}{2n} \geq \frac{N}{m} 
$$ 
winners to be deleted from~$B_0$ in the first round. 
Thus, there remain in $B_1$ at most $N\left(1 - \frac{1}{m}\right)$ 
elements after the first round, and, successively
applying this argument, $B_k$ contains at most 
$N\left(1 - \frac{1}{m}\right)^k$ elements after $k$ rounds.  Since
\mbox{$N\leq 2^m$} and the procedure terminates if 
$\|B_k\| \leq 2(n+1)$ for some
integer~$k$, it
suffices to show that some polynomial $k(m)$ of fixed degree satisfies
$$
\left( 1 - \frac{1}{m} \right)^{k(m)}\leq 2(n+1)2^{-m}.
$$  
This follows from the fact that $\lim\limits_{m\rightarrow \infty}\left(
\left( 1 - \frac{1}{m} \right)^{m^2} \right)^{m^{-1}} =e^{-1} < 
\frac{1}{2}$ 
implies that 
$\left( 1 - \frac{1}{m} \right)^{m^2} = {\cal O}(2^{-m})$.  As
in each round $n-1 < m$ strings of length $m$ are added to $A_m$, the
length of (the coding of) $A_m$ is indeed bounded above by some 
polynomial of degree~4.  

Note that the set
$$ 
C  \equalsdef  \left \{ \<x,a_{|x|}\>\
\begin{tabular*}{10cm}{|l}
\ $a_{|x|}$ is encoding of an advice $A_{|x|}$ and 
$x \in B_{k(|x|)}$, or $(\exists t_j)$ \\
\ $[t_j$ is a team of $A_{|x|}$ and $x$ belongs 
to or is yielded by $t_j ]$
\end{tabular*}
\right\}
$$
witnesses $L\in \ppoly$ (as stated in Theorem~\ref{thm:small-circuits}),
since clearly $C$ is a set in P and 
\mbox{$L = \{ x\, | \, \<x,a_{|x|}\>\in C \}$}.

\smallskip

Now we are ready to prove 
$L \in \mbox{\rm Low}_2$.  Let $D\in \np^{\protect\scriptsize
  \np^{L}}$ be witnessed by some NPOMs $N_1$ and~$N_2$, that is,
\mbox{$D=L(N_{1}^{L(N_{2}^L)})$}. Let $q(\ell)$ be a polynomial bound on the
length of all queries that can be asked in this computation on an
input of length $\ell$.  We describe below an NPOM $M$ and an \np\ 
oracle set~$E$ for which $D=L(M^E)$.

On input $x$, $M$ guesses for each length $m$, $1\leq m\leq q(|x|)$,
all possible polynomially length-bounded advice sets $A_m$ for
$L^{=m}$, simultaneously guessing witnesses (that is, an accepting
path of $N$ on input~$z$) that each string~$z$ in any guessed
advice set is in~$L^{=m}$. To check on each path 
whether the guessed sequence of
advice sets is correct, $M$ queries its oracle $E$ whether it contains
the string $\<x, A_1, \ldots , A_{q(|x|)}\>$, where
$$ 
E  \equalsdef  \left \{ \<x, A_1, \ldots , A_{q(|x|)}\> \
\begin{tabular*}{9.6cm}{|l}
\ $(\exists m : 1 \leq m \leq q(|x|))\, 
(\exists y_m : |y_m|=m )\, (\exists w_m)\, [ w_m$ \\
\ is an accepting path of $N(y_m)$, yet $y_m$ is neither \\
\ a string in $A_m$ nor is yielded by any team of $A_m]$
\end{tabular*}
\right\}
$$
is clearly a set in \np\@.  If the answer is ``yes,'' then some
guessed advice is incorrect, and $M$ rejects on that computation. If
the answer is ``no,'' then each guessed advice is correct for any
possible query of the respective length. Thus, $M$ now can simulate
the computation of $N_{1}^{L(N_{2})}$ on input $x$ using the selector
$f$ and the relevant advice $A_m$ to answer any question of $N_2$
correctly.  Hence, $D\in \np^{\protect\scriptsize \np}$.~\hspace*{\fill}$\Box $

\medskip

Ogihara has shown that if $\np \seq \mbox{\rm P-mc}(c\log n)$
for some $c < 1$, then \mbox{$\p = \np$}~\cite{ogi:j:comparable}\@.
Since by the proof of Theorem~\ref{thm:fair-s-p-mc},
$\mbox{\rm Fair-S}(c\log n, 1)$ is contained in $\mbox{\rm P-mc}(c\log n)$, 
$c < 1$, we have immediately the following corollary to Ogihara's result.
(Although Ogihara's result in~\cite{ogi:j:comparable} is also established 
for certain complexity classes other than NP,
we focus on the NP case only.)

\begin{corollary} \quad
  If $\np \seq \mbox{\rm Fair-S}(c\log n, 1)$ for some $c < 1$, then
  \mbox{$\p = \np$}.
\end{corollary}

\section[An Extended Selectivity Hierarchy Capturing 
Boo\-lean Closures of P-Sel]{An Extended Selectivity
  Hierarchy Capturing Boo\-lean Closures of P-Selective Sets}
\label{sec:gch}

\subsection{Distinguishing Between and Capturing Boo\-lean Closures
  of P-Selective Sets}
\label{subsec:capture}

Hemaspaandra and Jiang \cite{hem-jia:j:psel} noted that
the class P-Sel is closed under exactly those Boolean connectives that
are either completely degenerate or almost-completely degenerate.  In
particular, P-Sel is not closed under intersection or union, and is
not even closed under marked union (join). This raises the question of
how complex, e.g., the intersection of two P-selective sets is.  Also,
is the class of unions of two P-selective sets more or less complex
than the class of intersections of two P-selective sets?
Theorem~\ref{thm:boolean-pmc} establishes that, in terms of P-mc
classes, unions and intersections of sets in P-Sel are
indistinguishable (though they both are different from exclusive-or).
However, we will note as Theorem~\ref{thm:boolean-gc} that the GC
hierarchy (defined below) does distinguish between these classes, thus
capturing the closures of P-Sel under certain Boolean connectives
more tightly.

\begin{definition}
\label{def:gch}
\quad
Let~$g_1$,~$g_2$, and~$g_3$ be threshold functions.  

Define
$\mbox{\rm GC}(g_1(\cdot),g_2(\cdot),g_3(\cdot))$ to be the class 
of sets $L$ for
which there exists a polynomial-time computable function $f$ such that
for each $n\geq 1$ and any distinct input strings $y_1,\ldots ,y_n$,
\begin{enumerate}
\item
$f(y_1,\ldots ,y_n) \seq \{y_1,\ldots ,y_n\}$ and 
$\|f(y_1,\ldots ,y_n)\| \leq g_2(n)$, and

\item
$\| L \cap \{y_1,\ldots ,y_n\}\| \geq g_1(n) 
\Lora \| L \cap f(y_1,\ldots ,y_n)\| \geq g_3(n)$.
\end{enumerate}
\end{definition}

\begin{remark}
\label{rem:def-gch}
\begin{enumerate}
\item The notational conventions described
after Definition~\ref{def:S-fair-S} also apply to 
Definition~\ref{def:gch}.

\item
For constant thresholds $b$, $c$, $d$, we can equivalently (i.e.,
without changing the class) require in the definition that the
selector $f$ for a set $L$ in $\mbox{\rm GC}(b,c,d)$, on all input sets of
size at least $c$, must output {\em exactly\/} $c$ strings. This is
true because if $f$ outputs fewer than $c$ strings, we can define a
new selector $f'$ that outputs all strings output by $f$ and
additionally $\|f\| - c$ arbitrary input strings not output by $f$,
and $f'$ is still a $\mbox{\rm GC}(b,c,d)$-selector for $L$. This will be
useful in the proof of Lemma~\ref{lemma:gc:diag}.
\end{enumerate}
\end{remark}

The GC classes generalize the S classes of Section~\ref{sec:structure},
and as before, we also consider Fair-GC classes by additionally
requiring the ``fairness condition.''  Let GCH\label{ind:gch} 
denote $\bigcup_{i,j,k \geq 1}\mbox{\rm GC}(i,j,k)$.  
The internal structure of GCH will be analyzed
in Section~\ref{subsec:structure}.

A class ${\tweak\cal C} \seq 
{\cal P}(\sigmastar)$ of sets is said to be {\em nontrivial\/} 
if ${\tweak\cal C}$ contains infinite sets, 
but not all sets of strings over~$\Sigma$.  For example, the class
$\mbox{\rm Fair-GC}(\lceil \frac{n}{2} \rceil , \lceil \frac{n}{2} \rceil
, 1)$ equals ${\cal P}(\sigmastar)$ if $n$ is odd, and 
is therefore called trivial.  
First we note below that the largest nontrivial GC class,
$\mbox{\rm Fair-GC}( \lfloor \frac{n}{2} \rfloor , \lfloor \frac{n}{2}
\rfloor , 1)$, and thus all of GCH, is
contained in the P-mc hierarchy.

\begin{theorem}
\label{thm:gc-ceil}
\quad $\mbox{\rm Fair-GC}( \lfloor \frac{n}{2} \rfloor , \lfloor
\frac{n}{2} \rfloor , 1) \seq \mbox{\rm P-mc}(\mbox{poly})$.
\end{theorem}

\noindent
{\bf Proof.} \quad
Let $L \in \mbox{\rm Fair-GC}( \lfloor \frac{n}{2} \rfloor , \lfloor
\frac{n}{2} \rfloor , 1)$ via selector $f$. Fix any distinct inputs
$y_1,\ldots ,y_n$ such that $n\geq (\max\{|y_1|,\ldots ,|y_n|\})^2$.
Define a $\mbox{\rm P-mc}(n^2)$ function $g$ as follows: 
$g$ simulates $f(y_1,\ldots ,y_n)$ and outputs a ``0'' at each position
corresponding to an output string of~$f$, and outputs a ``1'' anywhere
else.  If all the strings having a ``1'' in the output of $g$
indeed are in~$L$, then at least one of the outputs of $f$ must be in~$L$,
since the ``fairness condition'' is met and $\| \{y_1,\ldots ,y_n \} \cap L\|
\geq \frac{n}{2}$. Thus, 
\[
(\chi_L(y_1), \ldots , \chi_L(y_n)) \neq
g(y_1,\ldots ,y_n),
\] 
and we have $L \in \mbox{\rm P-mc}(\mbox{poly})$ via~$g$.~\hfill$\Box$

\medskip

Now we state two lemmas that will be useful in the upcoming proofs of
Theorem~\ref{thm:boolean-pmc} and Theorem~\ref{thm:boolean-gc}.

\begin{lemma}
\label{lem:order}
{\rm \cite{buh-tor:j:pselective}}
\quad
Let $A \in \psel$ and $V \seq \sigmastar$.
The P-selector $f$ for $A$ induces a total order 
$\preceq_{f}$\label{ind:induced} on $V$ as follows:
For each $x$ and $y$ in $V$, define 
$x \preceq_{f} y$ if and only if
\[
(\exists u_1, \ldots , u_k)\, [x = u_1 \,\wedge\, y = u_k \,\wedge\, 
(\forall i : 2 \leq i \leq k)\, [f(u_{i-1}, u_{i}) = u_{i}]].
\]
Then, for all $x,y \in V$, 
\[
x \preceq_{f} y \Lolra (x \in A \Lora y \in A).
\]
\end{lemma}

The technique of constructing {\em widely-spaced\/} and {\em
  complexity-bounded} sets is a standard technique for constructing
P-selective sets.  This technique will be useful in the
diagonalization proofs of this section and will be applied in the form
presented
in~\cite{hem-jia:j:psel,hem-jia-rot-wat:c:multiselectivity,rot:thesis:promise}.
So let us first adopt some of the formalism used
in these papers.

Fix some wide-spacing function $\mu$ such that the spacing is at least
as wide as given by the following inductive definition: $\mu(0)=2$ and
\mbox{$\mu(i+1) = 2^{2^{\mu(i)}}$} for each~$i\ge 0$. Now define for
each~$k \geq 0$,
\[
R_k  \equalsdef  \{ i \,|\, i \in \N\, \wedge\,  \mu(k)\le i < \mu(k+1)\}, 
\]
and the following two classes of languages (where we will implicitly
use the standard correspondence between $\sigmastar$ and $\N$): 

\begin{eqnarray*}
{\cal C_1} & \equalsdef & \left\{ A \seq \N \ 
\begin{tabular*}{7.2cm}{|l}  %
\ $(\forall j\ge 0)\, 
[R_{2j}\cap A = \emptyset\,\wedge\, (\forall x,y \in R_{2j+1})\,$\\ 
\ $[(x\le y \,\wedge\, x\in A )\, \Longrightarrow  \,  y\in A]]$
\end{tabular*}
\right\} ; \\
{\cal C_2} & \equalsdef & \left\{ A \seq \N \ 
\begin{tabular*}{7.2cm}{|l}  %
\ $(\forall j\ge 0)\,
[R_{2j}\cap A = \emptyset\,\wedge\, (\forall x,y \in R_{2j+1})\,$\\
\ $[(x\le y \,\wedge\, y\in A) \, \Longrightarrow \, x\in A]]$
\end{tabular*}
\right\} .
\end{eqnarray*}

Then, the following lemma can be proven in the same vein as
in~\cite{hem-jia:j:psel}.

\begin{lemma}
{\rm \cite{hem-jia:j:psel}}
\label{lem:hemjia}
\quad
${\cal C_1} \cap \mbox{\rm E} \seq \psel\ $ and
$\ {\cal C_2} \cap \mbox{\rm E} \seq \psel$.
\end{lemma}

\begin{remark}
\label{rem:hj}
\begin{enumerate}
\item We will apply Lemma~\ref{lem:hemjia} in a slightly more general
  form in the proof of Theorem~\ref{thm:boolean-pmc} below. That is,
  in the definition of ${\cal C_1}$ and ${\cal C_2}$, the underlying
  ordering of the elements in the regions $R_{2j+1}$ need not be the
  standard lexicographical order of strings. We may allow {\em any\/}
  ordering $\prec$ that respects the lengths of strings and such that,
  given two strings, $x$ and $y$, of the same length, it can be
  decided in polynomial time whether $x \prec y$.

\item To accomplish the diagonalizations in this section, we need our
  enumeration of FP functions to satisfy a technical requirement.  Fix
  an enumeration of all polynomial-time transducers $\{ T_i\}_{i \geq
    1}$ having the property that each transducer appears infinitely
  often in the list.  That is, if $T = T_i$ (here, equality refers to
  the actual program) for some $i$, then there is an infinite set $J$
  of distinct integers such that for each $j \in J$, we have $T =
  T_j$. For each $k \geq 1$, let $f_k$ denote the function computed 
  by~$T_k$.  
  In the diagonalizations below, it is enough to diagonalize for all
  $k$ against some $T_{k'}$ such that $T_k = T_{k'}$, i.e., both
  compute~$f_k$.  In particular, for keeping the sets $L_1$ and $L_2$
  (to be defined in the upcoming proofs of
  Theorems~\ref{thm:boolean-pmc} and \ref{thm:boolean-gc}) in~E, we
  will construct $L_1$ and $L_2$ such that for all stages $j$ of the
  construction and for any set of inputs $X \seq R_{2j+1}$, the
  transducer computing $f_j(X)$ runs in time less than $2^{\max \{ |x|
    \, :\, x \in X \}}$ (i.e., the simulation of $T_j$ on input $X$ is
  aborted if it fails to be completed in this time bound, and the
  construction of $L_1$ and $L_2$ proceeds to the next stage). The
  diagonalization is still correct, since for each $T_i$ there is a
  number $b_i$ (depending only on $T_i$) such that for each $k \geq
  b_i$, if $T_i = T_k$, then for $T_k$ we will properly
  diagonalize---and thus $T_i$ is implicitly diagonalized against.
  
\item For each $j\geq 0$ and $k < \|R_{2j+1}\|$\label{ind:r2j}, let
  $r_{j,0}, \ldots , r_{j,k}$ denote the strings corresponding to the
  first $k+1$ numbers in region $R_{2j+1}$ (in the standard
  correspondence between $\sigmastar$ and $\N$). 
\end{enumerate}
\end{remark}

\begin{theorem}
\label{thm:boolean-pmc}
\begin{enumerate}
\item
\label{boolean-pmc:1}
$\psel \ \inter\  \psel \seq \mbox{\rm P-mc}(3)$, yet $\psel \ \inter\  \psel
\not\subseteq \mbox{\rm P-mc}(2)$.

\item
\label{boolean-pmc:2}
$\psel \ \union\  \psel \seq \mbox{\rm P-mc}(3)$, yet
$\psel \ \union\  \psel \not\subseteq \mbox{\rm P-mc}(2)$.

\item
\label{boolean-pmc:3}
$\psel \ \XOR\  \psel \not\subseteq \mbox{\rm P-mc}(3)$ and
$\psel \ \NXOR\  \psel \not\subseteq \mbox{\rm P-mc}(3)$.
\end{enumerate}
\end{theorem}

\noindent
{\bf Proof.} 
\ref{boolean-pmc:1}.\ \& \ref{boolean-pmc:2}.
Let $A \in \psel$ via $f$ and $B \in \psel$ via $g$, and let
$\preceq_{f}$ and $\preceq_{g}$ be the orders induced by
$f$ and $g$, respectively. Fix any inputs
$y_1$, $y_2$, and $y_3$ such that $y_1 \preceq_{f} y_2 \preceq_{f} y_3$.
Define a P-mc(3) function $h$ for $A\cap B$ as follows.
If $f$ and $g$ ``agree'' on any two of these strings (i.e., if there
exist $i,j \in \{1,2,3\}$ such that $i<j$ and $y_i \preceq_{g} y_j$),
then $h(y_1,y_2,y_3)$ outputs a ``1''
at position $i$ and a ``0'' at position~$j$. Otherwise (i.e., if 
\mbox{$y_3 \preceq_{g} y_2 \preceq_{g} y_1$}), define
$h(y_1,y_2,y_3)$ to output the string~101. In each case, we have
\[
(\chi_{A\cap B}(y_1), \chi_{A\cap B}(y_2), \chi_{A\cap B}(y_3)) \neq
h(y_1,y_2,y_3).
\] 
A similar construction works for $A \cup B$: Define
$h(y_1,y_2,y_3)$ to output the string 010 if 
\mbox{$y_3 \preceq_{g} y_2 \preceq_{g} y_1$}, and
as above in the other cases. This proves 
\mbox{$\psel \ \inter\  \psel \seq \mbox{\rm P-mc}(3)$} and
\mbox{$\psel \ \union\  \psel \seq \mbox{\rm P-mc}(3)$}.

\smallskip

For proving the diagonalizations, recall from the remark after 
Lemma~\ref{lem:hemjia} that
\mbox{$r_{j,0}, \ldots , r_{j,k}$} denote the smallest $k+1$ 
numbers in region $R_{2j+1}$. 
Define \mbox{$L_1  \equalsdef  \bigcup_{j\geq 0} L_{1,j}$}
and \mbox{$L_2  \equalsdef  \bigcup_{j\geq 0} L_{2,j}$}, where 
\[
L_{1,j}  \equalsdef  \left\{ i \in R_{2j+1} \
\begin{tabular*}{7cm}{|l}
\ $( f_j(r_{j,0} , r_{j,1}) \in \{ 00, 01 \} \,\wedge\, i \geq r_{j,1}) 
  \,\vee$\\
\ $( f_j(r_{j,0} , r_{j,1}) \in \{ 10, 11 \} \,\wedge\, i \geq r_{j,0})$
\end{tabular*}
\right\} ;
\]
\[
L_{2,j}  \equalsdef  \left\{ i \in R_{2j+1}  \
\begin{tabular*}{7cm}{|l}
\ $( f_j(r_{j,0} , r_{j,1}) \in \{ 00, 10 \} \,\wedge\, i \leq r_{j,0}) 
  \,\vee$\\
\ $( f_j(r_{j,0} , r_{j,1}) \in \{ 01, 11 \} \,\wedge\, 
i \leq r_{j,1})$
\end{tabular*}
\right\} .
\]
Clearly, by the above remark about the construction of $L_1$ and
$L_2$, we have that $L_1$ is in \mbox{${\cal C}_1 \cap \mbox{\rm E}$} 
and $L_2$ is in \mbox{${\cal C}_2 \cap \mbox{\rm E}$}.  
Thus, by Lemma~\ref{lem:hemjia}, $L_1$ and $L_2$ are in P-Sel.  
Supposing \mbox{$L_1 \cap L_2 \in \mbox{\rm P-mc}(2)$} via
$f_{j_{0}}$ for some~$j_{0}$, we have a string $f_{j_{0}}( r_{j_{0},0} ,
r_{j_{0},1})$ in $\{0,1\}^{2}$ that satisfies:
\[
(\chi_{L_1 \cap L_2}(r_{j_{0},0}), \chi_{L_1 \cap L_2}(r_{j_{0},1}))
\neq f_{j_{0}}( r_{j_{0},0} , r_{j_{0},1}).
\] 
However, in each of the
four cases for the membership of $r_{j_{0},0}$ and $r_{j_{0},1}$ in
$L_1 \cap L_2$, this is by definition of $L_1$ and $L_2$ exactly
what $f_{j_{0}}$ claims is impossible. Therefore, \mbox{$\psel \ \inter\  \psel
\not\subseteq \mbox{\rm P-mc}(2)$}.
Furthermore, since \psel\ is closed under complementation,
$\overline{L_1}$ and $\overline{L_2}$ are in P-Sel. Now assume
\mbox{$\psel \ \union\  \psel \subseteq \mbox{\rm P-mc}(2)$}. Then,
\mbox{$\overline{L_1} \cup \overline{L_2} = \overline{L_1 \cap L_2}$} is in 
$\mbox{\rm P-mc}(2)$, and since $\mbox{\rm P-mc}(2)$ is closed under complementation,
we have $L_1 \cap L_2 \in \mbox{\rm P-mc}(2)$, a contradiction.
Hence, \mbox{$\psel \ \union\  \psel \not\subseteq \mbox{\rm P-mc}(2)$}.

\smallskip

\ref{boolean-pmc:3}. 
Let $L_1  \equalsdef  \bigcup_{j\geq 0} L_{1,j}$, where $L_{1,j}$ is the set of 
all $i \in R_{2j+1}$ such that
\begin{description}
\item[(a)] $(f_j(r_{j,0} , r_{j,1} , r_{j,2}) \in \{ 100, 101 , 111\}
  \,\wedge\, i \geq r_{j,0})$ or

\item[(b)] $(f_j(r_{j,0} , r_{j,1} , r_{j,2}) = 011 \,\wedge\, i \geq
  r_{j,1})$ or

\item[(c)] $(f_j(r_{j,0} , r_{j,1} , r_{j,2}) \in \{ 001, 110 \}
  \,\wedge\, i \geq r_{j,2})$.
\end{description}
Thus, $L_1 \in {\cal C_1} \cap \mbox{\rm E}$, and by
Lemma~\ref{lem:hemjia}, $L_1 \in \psel$.  

For defining $L_2$, let us first
assume the following reordering of the elements in $R_{2j+1}$ for
each $j \geq 0$: \mbox{$r_{j,1} \prec r_{j,2} \prec r_{j,0} \prec r_{j,3}$}
and \mbox{$r_{j,s} \prec r_{j,s+1}$} if and only if 
\mbox{$r_{j,s} < r_{j,s+1}$} for $s
\geq 3$.  For any strings $x$ and $y$, we write $x \preceq y$ if $x
\prec y$ or $x = y$.  Now define $L_2  \equalsdef  \bigcup_{j\geq 0} L_{2,j}$,
where $L_{2,j}$ is the set of all $i \in R_{2j+1}$ such that
\begin{description}
\item[(a)] $(f_j(r_{j,0} , r_{j,1} , r_{j,2}) = 110 \,\wedge\, i
  \preceq r_{j,0})$ or

\item[(b)] $(f_j(r_{j,0} , r_{j,1} , r_{j,2}) \in \{ 010, 101 \}
  \,\wedge\, i \preceq r_{j,1})$ or

\item[(c)] $(f_j(r_{j,0} , r_{j,1} , r_{j,2}) = 100 \,\wedge\, i
  \preceq r_{j,2})$.
\end{description}
By Lemma~\ref{lem:hemjia} and the remark
following Lemma~\ref{lem:hemjia}, $L_2 \in \psel$.  
Note that for each \mbox{$j \geq 0$}, 
the set \mbox{$L_1 \cap R_{2j+1}$} is empty if $f_j(r_{j,0} , r_{j,1} ,
r_{j,2}) \in \{ 000, 010 \}$, and the set $L_2 \cap R_{2j+1}$ is empty
if $f_j(r_{j,0} , r_{j,1} , r_{j,2})$ is in $\{ 000, 001, 011, 111
\}$.  Now suppose $L_1 \xor L_2 \in \mbox{\rm P-mc}(3)$ via $f_{j_{0}}$
for some $j_{0}$, i.e., $f_{j_{0}}( r_{j_{0},0} , r_{j_{0},1} ,
r_{j_{0},2})$ is in $\{0,1\}^{3}$ and satisfies
\[
(\chi_{L_1 {\protect\scriptsize \xor} L_2}(r_{j_{0},0}), 
\chi_{L_1 {\protect\scriptsize \xor} L_2}(r_{j_{0},1}), 
\chi_{L_1 {\protect\scriptsize \xor} L_2}(r_{j_{0},2}))
\neq f_{j_{0}}( r_{j_{0},0} , r_{j_{0},1} , r_{j_{0},2}).
\] 
However, in each of the eight cases for the membership of 
$r_{j_{0},0}$, $r_{j_{0},1}$, and $r_{j_{0},2}$ in
$L_1 \xor L_2$, this is by definition of $L_1$ and $L_2$ exactly
what $f_{j_{0}}$ claims is impossible. Therefore, \mbox{$\psel \ \XOR\  \psel
\not\subseteq \mbox{\rm P-mc}(3)$}.
Since $L_1 \,\nxor\, \overline{L_2} = L_1 \xor L_2$ and 
$\overline{L_2} \in \psel$, this also implies that \mbox{$\psel \ \NXOR\  \psel
\not\subseteq \mbox{\rm P-mc}(3)$}.~\hfill$\Box$

\medskip

Note that Theorem~\ref{thm:boolean-pmc} does not contradict
Ogihara's result in~\cite{ogi:j:comparable} that %
${\Re}_{2\mbox{-}tt}^{p}(\psel)$ is contained in 
$\mbox{\rm P-mc}(2)$, since we
consider the union and intersection of two possibly {\em different\/}
sets in \psel, whereas the two queries in a
$\leq_{2\mbox{-}tt}^{p}$-reduction are asked to the {\em same\/} set
in \psel. Clearly, if \psel\ were closed under join, then we indeed
would have a contradiction.  However, \psel\ is not closed under join
\cite{hem-jia:j:psel}.

Next, we prove that in terms of the levels of the GCH hierarchy, the
class of intersections of P-selective sets can be clearly
distinguished from, e.g., the class of unions of P-selective sets.
This is in contrast with the P-mc hierarchy, which by the above
theorem is not refined enough to sense this distinction.  We note that
some parts of this Theorem~\ref{thm:boolean-gc} extend Hemaspaandra
and Jiang's results~\cite{hem-jia:j:psel}, and also
Rao's observation that $\psel\ \mbox{\bf op}\ \psel \not\seq
\mbox{\rm SH}$ for any Boolean operation {\bf op} chosen from 
$\{ \inter, \union, \XOR \}$~\cite{rao:perscomm:p-selectivity}.  Note
further that Part~\ref{boolean-gc:3} of Theorem~\ref{thm:boolean-gc}
still leaves a gap between the upper and the lower bound for
\mbox{$\psel \ \inter\ \psel$}.

\begin{theorem}~\label{thm:boolean-gc}~\begin{enumerate}
\item
\label{boolean-gc:1}
For each $k \geq 2$,
\begin{enumerate} 
\item ${\Oplus}_{k}(\psel) \seq \mbox{\rm GC}(1,k,1)$, but
  ${\Oplus}_{k}(\psel) \not\seq \mbox{\rm SH} \cup \mbox{\rm
    GC}(1,k-1,1)$, and

\item
  ${\union}_{k}(\psel) \seq \mbox{\rm GC}(1,k,1)$, but
  ${\union}_{k}(\psel) \not\seq \mbox{\rm SH} \cup \mbox{\rm
    GC}(1,k-1,1)$.
\end{enumerate}

\item
\label{boolean-gc:3}
$\psel \ \inter\  \psel \not\seq \mbox{\rm GC}(1,2,1)$, but for each 
integer-valued \fp\ function $k(0^n)$ 
satisfying \mbox{$1 \leq k(0^n) \leq n$}, 
$\psel \ \inter\  \psel \seq 
\mbox{\rm GC}( \lceil \frac{n}{k(0^n)} \rceil , k(0^n) , 1)$.

\item
\label{boolean-gc:4}
$\psel \ \mbox{\bf op}\  \psel \not\seq \mbox{\rm Fair-GC}(1,n-1,1)$ for
$\mbox{\bf op} \in \{ \inter, \XOR, \NXOR \}$.
\end{enumerate}
\end{theorem}

\noindent
{\bf Proof.} \ref{boolean-gc:1}.
Let $L = \oplus_k(A_1,  \ldots , A_k)$, where $A_i \in \psel$
via selector functions $s_i$ for $i \in \{1,\ldots , k\}$.
Let any inputs $x_1,\ldots , x_m$ be given, each having the
form $\underline{i}a$ for some $i \in \{1,\ldots , k\}$ and
$a\in \sigmastar$. For each $i$, play a knock-out tournament among
all strings $a$ for which $\underline{i}a$ belongs to the inputs,
where we say $a_1$ beats $a_2$ if $a_2 \preceq_{s_i} a_1$.
Let $w_1, \ldots , w_k$ be the winners of the $k$ tournaments.
Define a $\mbox{\rm GC}(1,k,1)$-selector for $L$ to output
$\{ \underline{1}w_1,\ldots , \underline{k}w_k \}$. Clearly, at least one of
these strings must be in $L$ if at least one of the inputs is in $L$.
The proof of ${\union}_{k}(\psel) \seq \mbox{\rm GC}(1,k,1)$ is 
similar. 

\smallskip

We only prove that $\psel \ \union\  \psel \not\seq \mbox{\rm SH}$ by uniformly
diagonalizing against all FP functions and all levels of SH\@. 
Define 
\begin{eqnarray*}
L_1  \equalsdef  \bigcup_{\<j,m\> \, :\, j \geq 0 \,\wedge\, 
m < \|R_{2j+1}\|} L_{1,\<j,m\>} & \mbox{ and } &
L_2  \equalsdef  \bigcup_{\<j,m\> \, :\, j \geq 0 \,\wedge\, 
m < \|R_{2j+1}\|} L_{2,\<j,m\>},
\end{eqnarray*}
where for each $j \geq 0$ and $m < \|R_{2j+1}\|$, the sets $L_{1,\<j,m\>}$
and $L_{2,\<j,m\>}$ are defined as follows:
\[
\left\{ i \in R_{2j+1} \,|\,
i > f_j(r_{j,0}, \ldots , r_{j,m}) \,\wedge\, 
f_j(r_{j,0}, \ldots , r_{j,m}) \in \{r_{j,0}, \ldots , r_{j,m}\}\right\};
\]
\[
\left\{ i \in R_{2j+1} \,|\,
i < f_j(r_{j,0}, \ldots , r_{j,m}) \,\wedge\, 
f_j(r_{j,0}, \ldots , r_{j,m}) \in \{r_{j,0}, \ldots , r_{j,m}\}\right\}.
\]
Clearly, $L_1 \in {\cal C}_1 \cap \mbox{\rm E}$ and $L_2 \in {\cal C}_2
\cap \mbox{\rm E}$.  Thus, by Lemma~\ref{lem:hemjia}, $L_1, L_2 \in
\psel$. Assume \mbox{$\psel \ \union\ \psel \seq \mbox{\rm SH}$}, and in
particular, $L_1 \cup L_2 \in \mbox{\rm S}(m_0)$ via $f_{j_0}$. If $m_0 <
\|R_{2j_0 + 1}\|$, then this contradicts the fact that
$f_{j_0}(r_{j_0,0}, \ldots , r_{j_0,m_0})$ selects a string not in $L_1
\cup L_2$ though $m_0$ of the inputs are in $L_1 \cup L_2$.  If $m_0 \geq
\|R_{2j_0 + 1}\|$, then by our assumption that each transducer $T_i$
appears infinitely often in the enumeration (see the remark after 
Lemma~\ref{lem:hemjia}),
there is an index $j_1$ such that $m_0 < \|R_{2j_1 + 1}\|$ and $T_{j_1}$
computes $f_{j_0}$, and thus $f_{j_0}$ is implicitly diagonalized
against.

\smallskip

\ref{boolean-gc:3}. Let $k(0^n)$ be a function as in the theorem.
Let $L = A \cap B$ for sets $A$ and $B$, where $A \in \psel$ via
$f$ and $B \in \psel$ via $g$. We will define a 
$\mbox{\rm GC}( \lceil \frac{n}{k(0^n)} \rceil , k(0^n) , 1)$-selector
$s$ for $L$. Given $n$ elements, rename them with respect to the
linear order induced by~$f$, i.e., we have
$x_1 \preceq_{f} x_2 \preceq_{f} \cdots \preceq_{f} x_n$. 
Let $k  \equalsdef  k(0^n)$. Now let $h$ be the unique permutation of
$\{ 1, \ldots , n\}$ such that for each $i,j \in \{ 1, \ldots , n\}$,
$h(i) = j$ if and only if $x_i$ is the $j$th element in the
linear ordering of $\{ x_1, \ldots , x_n\}$ induced by $g$.
Partition the set $\{ 1, \ldots , n\}$ into $k$ regions of at most
$\lceil \frac{n}{k} \rceil$ elements:
\begin{eqnarray*}
R(l) &  \equalsdef  & \left\{ 
(l-1)\left\lceil \frac{n}{k} \right\rceil + 1, 
(l-1)\left\lceil \frac{n}{k} \right\rceil + 2,
\ldots , l\left\lceil \frac{n}{k} \right\rceil \right\}
\ \ \mbox{for $1\leq l \leq k-1$, and}\\
R(k) &  \equalsdef  & \left\{ (k-1)\left\lceil \frac{n}{k} \right\rceil + 1, 
(k-1)\left\lceil \frac{n}{k} \right\rceil + 2, \ldots , n \right\}.
\end{eqnarray*}
Define $s(x_1, \ldots , x_n)  \equalsdef  \{ a_1, \ldots a_k \}$, where
$a_l  \equalsdef  x_{m(l)}$ and $m(l)$ is the $m \in R(l)$ such that $h(m)$ is
maximum. Thus, for each region $R(l)$, $a_l$ is the ``most likely'' element
of its region to belong to~$B$. Consider the permutation matrix of $h$
with elements $(i, h(i))$, for $1\leq i \leq n$. Let $c_A$ be the
``cutpoint'' for $A$ and let $c_B$ be the ``cutpoint'' for $B$, i.e., 
\[
\begin{array}{rcr}
\{x_i \, | \, i < c_A \} \seq \overline{A} & \mbox{ and } & 
\{x_i \, | \, i \geq c_A \} \seq A;\\
\{x_{h(i)} \, | \, {h(i)} < c_B \} \seq \overline{B} & \mbox{ and } & 
\{x_{h(i)}\, | \, {h(i)} \geq c_B \} \seq B.
\end{array}
\]
Define
\[
\begin{array}{lcllcl}
A_{\mbox{\protect\scriptsize out}} &  \equalsdef  & 
\{x_i \, | \, i < c_A \}; & 
\hspace*{1cm} A_{\mbox{\protect\scriptsize in}}  &  
\equalsdef  & \{x_i \, | \, i \geq c_A \};\\
B_{\mbox{\protect\scriptsize out}} &  \equalsdef  & 
\{x_{h(i)} \, | \, {h(i)} < c_B \}; & 
\hspace*{1cm} B_{\mbox{\protect\scriptsize in}}  &  
\equalsdef  & \{x_{h(i)}\, | \, {h(i)} \geq c_B \}.
\end{array}
\]
Since $A_{\mbox{\protect\scriptsize in}} \cap
B_{\mbox{\protect\scriptsize in}} \seq A \cap B$, it remains to show
that at least one of the outputs $a_l$ of
$s$ is in $A_{\mbox{\protect\scriptsize in}} \cap
B_{\mbox{\protect\scriptsize in}}$, if the promise 
\mbox{$\| \{ x_1, \ldots , x_n\} \cap L \| \geq \lceil
\frac{n}{k} \rceil$} is met.  First observe that for each~$l$,
if $i \geq c_A$ holds for each $i \in R(l)$ and $R(l)$ contains an
index $i_0$ such that $h(i_0) \geq c_B$, then $a_l \in
A_{\mbox{\protect\scriptsize in}} \cap B_{\mbox{\protect\scriptsize
    in}}$.  On the other hand, if $c_A$ ``cuts'' a region $R(l_0)$,
then in the worst case we have $a_{l_0} = (l_0-1)\lceil \frac{n}{k}
\rceil + 1$ and $c_A = (l_0-1)\lceil \frac{n}{k} \rceil + 2$, and thus
$a_{l_0} \not\in A_{\mbox{\protect\scriptsize in}}$ and at most
$\lceil \frac{n}{k} \rceil - 1$ elements of
$A_{\mbox{\protect\scriptsize in}}$ can have an index in~$R(l_0)$.
However, if $\| \{ x_1, \ldots , x_n\} \cap L \| \geq \lceil
\frac{n}{k} \rceil$, then there must exist an $l_1$ with $l_1 > l_0$
such that for each $i \in R(l_1)$ it holds that $i \geq c_A$, and
thus, $a_{l_1} \in A_{\mbox{\protect\scriptsize in}} \cap
B_{\mbox{\protect\scriptsize in}}$.  This proves $L \in \mbox{\rm GC}(
\lceil \frac{n}{k} \rceil , k , 1)$ via~$s$.

The proof of $\psel \ \inter\ \psel \not\seq \mbox{\rm GC}(1,2,1)$ is
similar as in Part~\ref{boolean-gc:4}.

\smallskip

\ref{boolean-gc:4}. We only prove 
$\psel \ \inter\ \psel \not\seq \mbox{\rm Fair-GC}(1,n-1,1)$ (the other
cases are similar). Define 
\[
L_1  \equalsdef  \left\{ i \
\begin{tabular*}{9.5cm}{|l}
\ $(\exists j \geq 0)\, [ i \in R_{2j+1}$ and $i \geq w_j $ for 
the smallest string  \\
\ $w_j \in R_{2j+1}$ such that $ f_j(R_{2j+1}) \seq R_{2j+1} - \{w_j\}]$ 
\end{tabular*}
\right\} ;
\]
\[
L_2  \equalsdef  \left\{ i \
\begin{tabular*}{9.5cm}{|l}
\ $(\exists j \geq 0)\, [ i \in R_{2j+1}$ and $i \leq w_j $ for 
the smallest string  \\
\ $w_j \in R_{2j+1}$ such that $ f_j(R_{2j+1}) \seq R_{2j+1} - \{w_j\}]$ 
\end{tabular*}
\right\} .  \] As before, $L_1, L_2 \in \psel$. Assume there is a
$\mbox{\rm Fair-GC}(1,n-1,1)$-selector $f_{j_0}$ for $L_1 \cap L_2$.
First observe that the ``fairness condition'' is satisfied if $f_{j_0}$ has
all strings from $R_{2j_{0}+1}$ as inputs, since $\| R_{2j_{0}+1} \| =
2^{2^{\mu(2j_{0}+1)}} - \mu(2j_{0}+1)$ and the length of the largest
string in $R_{2j_{0}+1}$ is at most $2^{\mu(2j_{0}+1)}$.  For the
$\mbox{\rm Fair-GC}(1,n-1,1)$-selector $f_{j_0}$, there must exist a
smallest string $w_{j_0}\in R_{2j_{0}+1}$ such that $f_{j_{0}}(R_{2j_{0}+1})$
is contained in $R_{2j_{0}+1} - \{w_{j_{0}}\}$, and thus, 
$\{w_{j_0}\} = L_1 \cap
L_2 \cap R_{2j_{0}+1}$. This would contradict
$f_{j_{0}}(R_{2j_{0}+1})$ not selecting~$w_{j_0}$.
\hspace*{\fill}$\Box$

\medskip

Statement~\ref{boolean-gc:3} of the above theorem immediately gives
the first part of Corollary~\ref{cor:boolean-gc:3}.  Note that, even
though this $\mbox{\rm GC}( \sqrt{n} , \sqrt{n} , 1)$ upper bound on
$\psel \ \inter\ \psel$ may not be strong enough to prove the second
part of the corollary, the proof of this second part does easily
follow from the $\psel \ \inter\ \psel \seq \mbox{\rm P-mc}(3)$ result of
Theorem~\ref{thm:boolean-pmc} via Ogihara's result that the assumption
$\np \seq \mbox{\rm P-mc}(3)$ implies the collapse of $\p = \np$
\cite{ogi:j:comparable}.

\begin{corollary}
\label{cor:boolean-gc:3}
\begin{enumerate}
\item
$\psel \ \inter\  \psel \seq 
\mbox{\rm GC}( \sqrt{n} , \sqrt{n} , 1)$.

\item
$\np \seq \psel \ \inter\  \psel \Lora \p = \np$.
\end{enumerate}
\end{corollary}

\subsection{The Structure of the GC Hierarchy}
\label{subsec:structure}

In this subsection, we study the internal structure of~GCH\@.  We
start with determining for which parameters~$b$,~$c$, and~$d$ the
class $\mbox{\rm GC}(b,c,d)$ is nontrivial 
(i.e., satisfies \mbox{$\mbox{\rm GC}(b,c,d) \neq
{\cal P}(\sigmastar)$}, yet contains not only finite sets).  Recall
that $w_{i,1}, \ldots , w_{i,s}$ are the lexicographically smallest
$s$ length $e(i)$ strings, for $i\geq 0$ and $s \leq 2^{e(i)}$
(the function $e(i)$ is defined in Section~\ref{sec:structure}).  The
proofs of some of the more technical lemmas in this subsection are
deferred to Section~\ref{subsec:hard-proofs}. For instance, the
proof of Lemma~\ref{lemma:gc:trivial} below can be found in 
Section~\ref{subsec:hard-proofs}.

\begin{lemma}
\label{lemma:gc:trivial}
\quad
Let $b,c,d \in \N^+$ with $d \leq c$ and $d \leq b$. Then, 
\begin{enumerate}
\item 
\label{lem:gc:trivial:1}
$(\exists A)\, [A \in \mbox{\rm GC}(b,c,d)\ \wedge\ \| A\| = \infty]$, and

\item  
\label{lem:gc:trivial:2}
$(\exists B)\, [B \not\in \mbox{\rm GC}(b,c,d)\ \wedge\ \| B\| = \infty]$.
\end{enumerate}
\end{lemma}

\begin{theorem}
\label{thm:gc:trivial}
\quad
Let $b,c,d \in \N^+$.
\begin{enumerate}
\item 
\label{thm:gc:trivial:1}
Every set in $\mbox{\rm GC}(b,c,d)$ is finite if and only if 
$d > b$ or $d > c$.

\item 
\label{thm:gc:trivial:2}
If $d \leq b$ and $d \leq c$, then $\mbox{\rm GC}(b,c,d)$ is nontrivial.
\end{enumerate}
\end{theorem}

\noindent
{\bf Proof.}
\quad
If $d>c$ or $d>b$, then by Definition~\ref{def:gch}, every set in
$\mbox{\rm GC}(b,c,d)$ is finite.  On the other hand, if $d \leq b$ and $d
\leq c$, then by Lemma~\ref{lemma:gc:trivial}.\ref{lem:gc:trivial:1},
there is an infinite set in $\mbox{\rm GC}(b,c,d)$.  Hence, every set in
$\mbox{\rm GC}(b,c,d)$ is finite if and only if $d > b$ or $d > c$. 
Furthermore, if $d \leq b$ and $d \leq c$, then
$\mbox{\rm GC}(b,c,d) \neq {\cal P}(\sigmastar)$ by
Lemma~\ref{lemma:gc:trivial}.\ref{lem:gc:trivial:2}.~\hfill$\Box$

\medskip

Now we turn to the relationships between the nontrivial classes
within GCH\@. Given any parameters $b,c,d$ and $i,j,k$, we seek to
determine which of $\mbox{\rm GC}(b,c,d)$ and $\mbox{\rm GC}(i,j,k)$ is
contained in the other class (and if this inclusion is strict), or
whether they are mutually incomparable. For classes $\cal A$ and $\cal
B$, let ${\cal A}\bowtie {\cal B}$\label{ind:bowtie} 
denote that $\cal A$ and $\cal B$
are incomparable, i.e., ${\cal A} \not\seq {\cal B}$ and ${\cal B}
\not\seq {\cal A}$.  Theorem~\ref{thm:structure-gc} will establish
these relations for almost all the cases and is proven by making
extensive use of the Inclusion Lemma
and the Diagonalization Lemma 
below. The proofs of Lemmas~\ref{lemma:gc:inclusion}
and~\ref{lemma:gc:diag} can be found in Section~\ref{subsec:hard-proofs}.

\begin{lemma}[(Inclusion Lemma)]
\label{lemma:gc:inclusion}
{\rm
\quad
Let $b, c, d \in \N^+$ and $l, m, n \in \N$ be given such that each
GC class below is nontrivial. Then,
\begin{enumerate}
\item  
\label{gc:inclusion:1}
$\mbox{\rm GC}(b,c,c)=\mbox{\rm S}(b,c)$.

\item  
\label{gc:inclusion:2}
$\mbox{\rm GC}(b,c,d+n) \seq \mbox{\rm GC}(b+l,c+m,d)$.

\item  
\label{gc:inclusion:3}
If $l\ge n$ and $m\ge n$, then 
$\mbox{\rm GC}(b,c,c) \seq \mbox{\rm GC}(b+l,c+m,c+n)$.%

\item  
\label{gc:inclusion:4}
If $l\le n$ and $m\le n$, then 
$\mbox{\rm GC}(b+l,c+m,d+n) \seq \mbox{\rm GC}(b,c,d)$.
\end{enumerate}
}
\end{lemma}

\begin{lemma}[(Diagonalization Lemma)]
{\rm
\label{lemma:gc:diag}
\quad
Let $b, c, d \in \N^+$ and $l, m, n, q \in \N$ be given such that each
GC class below is nontrivial. Then,
\begin{enumerate}
\item 
\label{gc:diag:1}
If $l\ge n+1$, then
$(\exists L)\, [L\in \mbox{\rm GC}(b+l,c+m,d+n)-\mbox{\rm GC}(b,c+q,d)]$.

\item 
\label{gc:diag:2}
If $m\ge n+1$, then
$(\exists L)\, [L\in \mbox{\rm GC}(b+l,c+m,d+n)-\mbox{\rm GC}(b+q,c,d)]$.

\item 
\label{gc:diag:3}
If ($n\ge l+1$ or $n\ge m+1$), then 
$(\exists L)\, [L\in \mbox{\rm GC}(b,c,d)-\mbox{\rm GC}(b+l,c+m,d+n)]$.
\end{enumerate}
}
\end{lemma}

\begin{theorem}
\label{thm:structure-gc}
\quad
Let $b, c, d \in \N^+$ and $i, j, k \in \N$ be given such that each GC
class below is nontrivial.  Then,
\begin{enumerate}
\item 
\label{structure-gc:1}
$\mbox{\rm GC}(b,c,d+k) \subset \mbox{\rm GC}(b+i,c+j,d)$
if $i \geq 1$ or $j \geq 1$ or $k \geq 1$.

\item
\label{structure-gc:2}
$\mbox{\rm GC}(b,c+j,d+k) \subset \mbox{\rm GC}(b+i,c,d)$ if $1 \leq j \leq k$.

\item 
\label{structure-gc:3}
$\mbox{\rm GC}(b,c+j,d+k) \bowtie \mbox{\rm GC}(b+i,c,d)$ if $j > k \geq 1$.

\item 
\label{structure-gc:4}
$\mbox{\rm GC}(b+i,c,d+k) \subset \mbox{\rm GC}(b,c+j,d)$ if $1 \leq i \leq k$.
 
\item 
\label{structure-gc:5}
$\mbox{\rm GC}(b+i,c,d+k) \bowtie \mbox{\rm GC}(b,c+j,d)$ if $i > k \geq 1$.

\item 
\label{structure-gc:6}
$\mbox{\rm GC}(b+i,c,d) \bowtie \mbox{\rm GC}(b,c+j,d)$ if $i \geq 1$ and $j \geq 1$.

\item 
\label{structure-gc:7}
$\mbox{\rm GC}(b+i,c+j,d+k) \subset \mbox{\rm GC}(b,c,d)$ if ($1 \leq i < k$
and $1 \leq j \leq k$) or ($1 \leq j < k$ and $1 \leq i \leq k$).
 
\item
\label{structure-gc:8}
$\mbox{\rm GC}(b+i,c+j,d+k) = \mbox{\rm GC}(b,c,d)$ if $i = j = k$ and $c = d$.
 
\item
\label{structure-gc:9}
$\mbox{\rm GC}(b+i,c+j,d+k) \bowtie \mbox{\rm GC}(b,c,d)$ if $1 \leq i < k < j$ or 
$1 \leq j < k < i$.  
\end{enumerate}
\end{theorem}

\noindent
{\bf Proof.} 
\quad
The proof is done by repeatedly applying
Lemma~\ref{lemma:gc:inclusion} and Lemma~\ref{lemma:gc:diag}. Unless
otherwise specified, $l$, $m$, and $n$ in the lemmas correspond to
$i$, $j$, and $k$ in this proof.

\smallskip

\ref{structure-gc:1}.  The inclusion is clear (see
Lemma~\ref{lemma:gc:inclusion}.\ref{gc:inclusion:2}).  For the
strictness of the inclusion, we have to consider three cases.  If
$i\geq 1$, then by Lemma~\ref{lemma:gc:diag}.\ref{gc:diag:1} with $n =
q = 0$, there exists a set $L \in \mbox{\rm GC}(b+i,c+j,d) -
\mbox{\rm GC}(b,c,d)$. By
Lemma~\ref{lemma:gc:inclusion}.\ref{gc:inclusion:2} with $l = m = 0$,
$L \not\in \mbox{\rm GC}(b,c,d+k)$.  The case of $j\geq 1$ is treated
similar, using Lemma~\ref{lemma:gc:diag}.\ref{gc:diag:2} instead of
Lemma~\ref{lemma:gc:diag}.\ref{gc:diag:1}.  Finally, if $k\geq 1$,
then by Lemma~\ref{lemma:gc:diag}.\ref{gc:diag:3} with $l = m = 0$, we
have $L \in \mbox{\rm GC}(b,c,d) - \mbox{\rm GC}(b,c,d+k)$. By
Lemma~\ref{lemma:gc:inclusion}.\ref{gc:inclusion:2} with $n = 0$, $L
\in \mbox{\rm GC}(b+i,c+j,d)$.

\smallskip

\ref{structure-gc:2}.  Applying
Lemma~\ref{lemma:gc:inclusion}.\ref{gc:inclusion:4} with $l = 0$ and
then Lemma~\ref{lemma:gc:inclusion}.\ref{gc:inclusion:2} with $m = n =
0$, we have $\mbox{\rm GC}(b,c+j,d+k) \seq \mbox{\rm GC}(b,c,d) \seq
\mbox{\rm GC}(b+i,c,d)$.  By Lemma~\ref{lemma:gc:diag}.\ref{gc:diag:3}
with $l = 0$ (i.e., $n\geq 1$), there exists a set $L \in
\mbox{\rm GC}(b,c,d) - \mbox{\rm GC}(b,c+j,d+k)$. By
Lemma~\ref{lemma:gc:inclusion}.\ref{gc:inclusion:2} with $m = n = 0$,
$L \in \mbox{\rm GC}(b+i,c,d)$.

\smallskip

\ref{structure-gc:3}.  ``$\not\seq$'' follows from 
Lemma~\ref{lemma:gc:diag}.\ref{gc:diag:2} with $q = i$ and $l = 0$.
``$\not\supseteq$'' follows as in Part~\ref{structure-gc:2}.  

\smallskip

\ref{structure-gc:4}.  Applying
Lemma~\ref{lemma:gc:inclusion}.\ref{gc:inclusion:4} with $m = 0$ and
then Lemma~\ref{lemma:gc:inclusion}.\ref{gc:inclusion:2} with $l = n =
0$, we have $\mbox{\rm GC}(b+i,c,d+k) \seq \mbox{\rm GC}(b,c,d) \seq
\mbox{\rm GC}(b,c+m,d)$.  The strictness of the inclusion follows as in
Part~\ref{structure-gc:2}, where
Lemma~\ref{lemma:gc:diag}.\ref{gc:diag:3} is applied with $m = 0$
instead of $l = 0$.

\smallskip

\ref{structure-gc:5}.  ``$\not\seq$'' follows from
Lemma~\ref{lemma:gc:diag}.\ref{gc:diag:1} with $q = j$ and $m = 0$.
``$\not\supseteq$'' holds by Lemma~\ref{lemma:gc:diag}.\ref{gc:diag:3}
with $m = 0$ (i.e., $n\geq 1$) and
Lemma~\ref{lemma:gc:inclusion}.\ref{gc:inclusion:2} with $l = n = 0$.

\smallskip

\ref{structure-gc:6}.  ``$\not\seq$'' holds, as by
Lemma~\ref{lemma:gc:diag}.\ref{gc:diag:1} with $q = j$ and $m = n =0$, 
there exists a set $L$ in \mbox{$\mbox{\rm GC}(b+i,c,d) - \mbox{\rm GC}(b,c+j,d)$}.
``$\not\supseteq$'' similarly follows from 
Lemma~\ref{lemma:gc:diag}.\ref{gc:diag:2} with $q = i$ and $l = n =0$.

\smallskip

\ref{structure-gc:7}.  By
Lemma~\ref{lemma:gc:inclusion}.\ref{gc:inclusion:4},
$\mbox{\rm GC}(b+i,c+j,d+k) \subseteq \mbox{\rm GC}(b,c,d)$.
By Lemma~\ref{lemma:gc:diag}.\ref{gc:diag:3}, if $n > l$ or $n > m$,
then there exists a set $L \in \mbox{\rm GC}(b,c,d) - \mbox{\rm GC}(b+i,c+j,d+k)$.

\smallskip

\ref{structure-gc:8}.  The equality follows from
Lemma~\ref{lemma:gc:inclusion}.\ref{gc:inclusion:3} and
Lemma~\ref{lemma:gc:inclusion}.\ref{gc:inclusion:4}.

\smallskip

\ref{structure-gc:9}.  Let $i < k < j$. Then, by
Lemma~\ref{lemma:gc:diag}.\ref{gc:diag:2} with $q = 0$, there exists a
set $L$ in \mbox{$\mbox{\rm GC}(b+i,c+j,d+k) - \mbox{\rm GC}(b,c,d)$}. Conversely, by
Lemma~\ref{lemma:gc:diag}.\ref{gc:diag:3}, there exists a set $L$ in 
\mbox{$\mbox{\rm GC}(b,c,d) - \mbox{\rm GC}(b+i,c+j,d+k)$}. If $j < k < i$, the
incomparability of $\mbox{\rm GC}(b,c,d)$ and $\mbox{\rm GC}(b+i,c+j,d+k)$
similarly follows from Lemma~\ref{lemma:gc:diag}.\ref{gc:diag:1} and
Lemma~\ref{lemma:gc:diag}.\ref{gc:diag:3}.
\hspace*{\fill}$\Box$

\medskip

\begin{figure}[tbp]
\centerline{\psfig{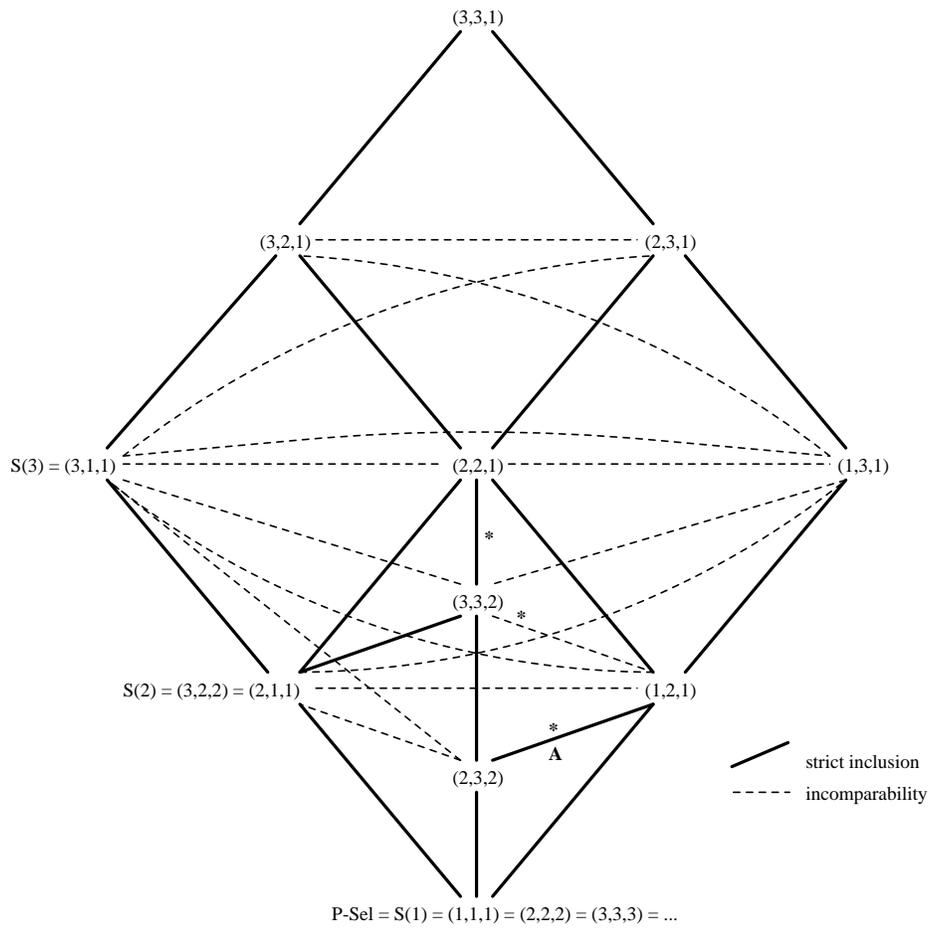}}
\caption{Relations between all nontrivial classes $\mbox{\rm GC}(b,c,d)$ 
with $1 \leq b,c,d \leq 3$.
\label{f:gc}}
\end{figure}

Note that Theorem~\ref{thm:structure-gc} does not settle all possible
relations between the GC classes. That is, the relation between
$\mbox{\rm GC}(b,c,d)$ and $\mbox{\rm GC}(b+i,c+j,d+k)$ is left open for the
case of ($k\leq i$ and $k\leq j$ and $c \neq d$).  Figure~\ref{f:gc}
shows the relations amongst all nontrivial classes $\mbox{\rm GC}(b,c,d)$
with $1 \leq b,c,d \leq 3$, as they are proven in
Theorem~\ref{thm:structure-gc} and Theorem~\ref{thm:nick} (those
relations not established by Theorem~\ref{thm:structure-gc} are marked
by ``$*$'' in Figure~\ref{f:gc} and are proven separately as
Theorem~\ref{thm:nick} below). For instance, \mbox{$\mbox{\rm S}(2) =
\mbox{\rm GC}(3,2,2) \subset \mbox{\rm GC}(3,3,2)$} holds by the first part of
Theorem~\ref{thm:structure-gc} with $b = 3$, \mbox{$c = d = 2$}, $i = k = 0$, 
and $j = 1$. The
``A'' in Figure~\ref{f:gc} indicates that, while the inclusion holds
by Lemma~\ref{lemma:gc:inclusion}.\ref{gc:inclusion:4}, the 
strictness of the inclusion was observed by A.~Nickelsen and
appears here with his kind permission.

\begin{theorem} 
\label{thm:nick}
\begin{enumerate}
\item \label{thm:nick:1} {\rm \cite{nic:perscomm:1994}} \quad $\mbox{\rm GC}(2,3,2) \subset
  \mbox{\rm GC}(1,2,1)$.

\item \label{thm:nick:2} $\mbox{\rm GC}(3,3,2) \bowtie \mbox{\rm GC}(1,2,1)$.

\item \label{thm:nick:3} $\mbox{\rm GC}(3,3,2) \subset \mbox{\rm GC}(2,2,1)$.
\end{enumerate}
\end{theorem}

\noindent
{\bf Proof.}
\quad
Both the inclusion $\mbox{\rm GC}(2,3,2) \seq \mbox{\rm GC}(1,2,1)$ and
the inclusion $\mbox{\rm GC}(3,3,2) \seq \mbox{\rm GC}(2,2,1)$ follow from
Lemma~\ref{lemma:gc:inclusion}.\ref{gc:inclusion:4} with $l = m = n =
1$.  We now provide those diagonalizations required to complete
the proof of the theorem.

\smallskip

\ref{thm:nick:1}.  For proving $\mbox{\rm GC}(1,2,1) \not\seq
\mbox{\rm GC}(2,3,2)$, we will define a set $L = \bigcup_{i \geq 1} L_i$
such that for each~$i$, $L_i \seq W_{i,4}$, and if $f_i(W_{i,4}) \seq
W_{i,4}$ and $\|f_i(W_{i,4})\| = 3$, then we make sure that
$\|L_i\| = 2$ and $\|L_i \cap f_i(W_{i,4})\| = 1$. This ensures that
for no $i\geq 1$ can $f_i$ be a $\mbox{\rm GC}(2,3,2)$-selector for $L$.
For example, this can be accomplished by defining $L_i$ as follows:
\begin{eqnarray*}
\chi_{L}(w_{i,1},\ldots ,w_{i,4}) = 0101 & \mbox{if} &
f_i(W_{i,4}) = \{ w_{i,1}, w_{i,2}, w_{i,3} \}, \\
\chi_{L}(w_{i,1},\ldots ,w_{i,4}) = 1010 & \mbox{if} &
f_i(W_{i,4}) = \{ w_{i,1}, w_{i,2}, w_{i,4} \}, \\
\chi_{L}(w_{i,1},\ldots ,w_{i,4}) = 1100 & \mbox{if} &
f_i(W_{i,4}) = \{ w_{i,1}, w_{i,3}, w_{i,4} \}, \\
\chi_{L}(w_{i,1},\ldots ,w_{i,4}) = 1100 & \mbox{if} &
f_i(W_{i,4}) = \{ w_{i,2}, w_{i,3}, w_{i,4} \}.
\end{eqnarray*}
Note that if $f_i(W_{i,4})$ outputs a string not in $W_{i,4}$ or the
number of output strings is different from 3, then (by
Definition~\ref{def:gch} and the remark following
Definition~\ref{def:gch}) $f_i$
immediately is disqualified from being a $\mbox{\rm GC}(2,3,2)$-selector for
$L$ (and we set $L_i = \emptyset$ in this case).  Thus, \mbox{$L \not\in
\mbox{\rm GC}(2,3,2)$}. On the other hand, $L \in \mbox{\rm GC}(1,2,1)$ can be
seen as follows: Given any set of inputs $X$ with $\|X\| \geq 2$, we
can w.l.o.g.\ assume that $X \seq \bigcup_{i\geq1} W_{i,4}$; since
smaller strings can be solved by brute force, we may even assume that
$X \seq W_{j,4}$ for some $j$. Suppose further that $\|L \cap X\| \geq
1$. Define $g(X)  \equalsdef  X$ if $\|X\| = 2$; and if $\|X\| > 2$, define
$g(X)$ to output $\{w_{j,1}, w_{j,4}\}$ if $\{w_{j,1}, w_{j,4}\} \seq
X$, and to output $\{w_{j,2}, w_{j,3}\}$ otherwise. Since \mbox{$\|L \cap
\{w_{j,1}, w_{j,4}\} \| = 1$} and \mbox{$\|L \cap \{w_{j,2}, w_{j,3}\} \| =
1$} holds in each of the four cases above, it follows that 
$\|L \cap g(X)\| \geq 1$. Hence, $L \in \mbox{\rm GC}(1,2,1)$ via~$g$.

\smallskip

\ref{thm:nick:2}.  For proving $\mbox{\rm GC}(1,2,1) \not\seq
\mbox{\rm GC}(3,3,2)$, $L$ is defined as $\bigcup_{i \geq 1} L_i$, where
\mbox{$L_i \seq W_{i,5}$}, and if $f_i(W_{i,5}) \seq W_{i,5}$ and
$\|f_i(W_{i,5})\| = 3$, then we make sure that $\|L_i\| = 3$ and
$\|L_i \cap f_i(W_{i,5})\| = 1$. This ensures that for no $i\geq 1$ can
$f_i$ be a $\mbox{\rm GC}(3,3,2)$-selector for $L$. For example, this can
be achieved by defining $L_i$ as follows:
\begin{eqnarray*}
\chi_{L}(w_{i,1},\ldots ,w_{i,5}) = 01011 & \mbox{if} &
f_i(W_{i,5}) = \{ w_{i,1}, w_{i,2}, w_{i,3} \}, \\
\chi_{L}(w_{i,1},\ldots ,w_{i,5}) = 10101 & \mbox{if} &
f_i(W_{i,5}) = \{ w_{i,1}, w_{i,2}, w_{i,4} \}, \\
\chi_{L}(w_{i,1},\ldots ,w_{i,5}) = 10110 & \mbox{if} &
f_i(W_{i,5}) = \{ w_{i,1}, w_{i,2}, w_{i,5} \}, \\
\chi_{L}(w_{i,1},\ldots ,w_{i,5}) = 01101 & \mbox{if} &
f_i(W_{i,5}) = \{ w_{i,1}, w_{i,3}, w_{i,4} \}, \\
\chi_{L}(w_{i,1},\ldots ,w_{i,5}) = 01011 & \mbox{if} &
f_i(W_{i,5}) = \{ w_{i,1}, w_{i,3}, w_{i,5} \}, \\
\chi_{L}(w_{i,1},\ldots ,w_{i,5}) = 01101 & \mbox{if} &
f_i(W_{i,5}) = \{ w_{i,1}, w_{i,4}, w_{i,5} \}, \\
\chi_{L}(w_{i,1},\ldots ,w_{i,5}) = 10101 & \mbox{if} &
f_i(W_{i,5}) = \{ w_{i,2}, w_{i,3}, w_{i,4} \}, \\
\chi_{L}(w_{i,1},\ldots ,w_{i,5}) = 11010 & \mbox{if} &
f_i(W_{i,5}) = \{ w_{i,2}, w_{i,3}, w_{i,5} \}, \\
\chi_{L}(w_{i,1},\ldots ,w_{i,5}) = 10110 & \mbox{if} &
f_i(W_{i,5}) = \{ w_{i,2}, w_{i,4}, w_{i,5} \}, \\
\chi_{L}(w_{i,1},\ldots ,w_{i,5}) = 11010 & \mbox{if} &
f_i(W_{i,5}) = \{ w_{i,3}, w_{i,4}, w_{i,5} \}.
\end{eqnarray*}
As argued above, this shows that $L \not\in \mbox{\rm GC}(3,3,2)$.  For
proving that $L$ is in $\mbox{\rm GC}(1,2,1)$, let a set $X$ of inputs be
given and suppose w.l.o.g.\ that $\|X\| \geq 3$ and $X \seq W_{j,5}$
for some $j$. Note that for each choice of $X$, at least one of
$\{w_{j,1}, w_{j,2}\}$, $\{w_{j,2}, w_{j,3}\}$, $\{w_{j,3},
w_{j,4}\}$, $\{w_{j,4}, w_{j,5}\}$, or $\{w_{j,5}, w_{j,1}\}$ must be
contained in $X$\@. On the other hand, each of $\{w_{j,1}, w_{j,2}\}$,
$\{w_{j,2}, w_{j,3}\}$, $\{w_{j,3}, w_{j,4}\}$, $\{w_{j,4},
w_{j,5}\}$, and $\{w_{j,5}, w_{j,1}\}$ has (by construction of $L$) at
least one string in common with $L_j$ if $L_j$ is not set to the empty
set. From these comments the action of the $\mbox{\rm GC}(1,2,1)$-selector
is clear.  

\smallskip

For proving $\mbox{\rm GC}(3,3,2) \not\seq \mbox{\rm GC}(1,2,1)$, define a set
$L \seq \bigcup_{i \geq 1} W_{i,3}$ as follows:
\begin{eqnarray*}
\chi_{L}(w_{i,1}, w_{i,2}, w_{i,3}) = 100 & \mbox{if} &
f_i(W_{i,3}) = \{ w_{i,2}, w_{i,3} \}, \\
\chi_{L}(w_{i,1}, w_{i,2}, w_{i,3}) = 010 & \mbox{if} &
f_i(W_{i,3}) = \{ w_{i,1}, w_{i,3} \}, \\
\chi_{L}(w_{i,1}, w_{i,2}, w_{i,3}) = 001 & \mbox{if} &
f_i(W_{i,3}) = \{ w_{i,1}, w_{i,2} \}.
\end{eqnarray*}
Since in each case $\|L \cap W_{i,3}\| = 1$ but $L \cap f_i(W_{i,3}) =
\emptyset$, $L$ cannot be in~$\mbox{\rm GC}(1,2,1)$. On the other
hand, $L$ is easily seen to be in $\mbox{\rm GC}(3,3,2)$ via a selector
that first solves all ``small'' inputs (i.e., those strings not of
maximum length) by brute force and then outputs two small members of
$L$ (and one arbitrary input) if those can be found, or three
arbitrary inputs if no more than one small member of $L$ is found by
brute force. Note that the $\mbox{\rm GC}(3,3,2)$-promise is not satisfied
in the latter case.

\smallskip

Part~\ref{thm:nick:3} follows from Part~\ref{thm:nick:2}, as
$\mbox{\rm GC}(1,2,1) \subset \mbox{\rm GC}(2,2,1)$.~\hfill$\Box$

\subsection{Some Proofs Deferred from Section~\ref{subsec:structure}}
\label{subsec:hard-proofs}

\noindent
{\bf Proof of Lemma~\ref{lemma:gc:trivial}.} 
\label{proof:gc:trivial}
\ref{lem:gc:trivial:1}. 
Let $A = \sigmastar$. Given $n$ distinct strings 
$y_1,  \ldots , y_n$, define 
\[
f(y_1, \ldots ,y_n) \equalsdef \left\{ 
\begin{array}{ll}
\{y_1, \ldots ,y_c\} & \mbox{if $n\geq c$}\\
\{y_1, \ldots ,y_n\} & \mbox{if $n< c$.}
\end{array}
\right.
\]
Clearly, $f\in \fp$, $f(y_1, \ldots ,y_n) \subseteq A$, and $\|f(y_1,
\ldots ,y_n)\|\leq c$.  

If \mbox{$\|\{y_1, \ldots , y_n\} \cap A\|\geq b$},
then $n\geq b$, and thus we have 
$$\|f(y_1, \ldots ,y_n) \cap A\| = c \geq d$$ 
if $n\geq c$, and if $n < c$, then
\[
\|f(y_1, \ldots ,y_n) \cap A\| = n
\geq b \geq d.
\]  
By Definition~\ref{def:gch}, $A \in \mbox{\rm GC}(b,c,d)$.
 
\smallskip

\ref{lem:gc:trivial:2}.  We will define $B \equalsdef \bigcup_{i \geq 1} B_i$
such that for no $i$ with $b+c-d+1 \leq 2^{e(i)}$ can $f_i$ be a
$\mbox{\rm GC}(b,c,d)$-selector for $B$. By our assumption about the
enumeration of FP functions (recall the remark after 
Lemma~\ref{lem:hemjia}), this suffices. For
each $i$ with 
$$
b+c-d+1 > 2^{e(i)},
$$ 
set $B_i \equalsdef \emptyset$.  
For each $i$ such that 
$$
b+c-d+1 \leq 2^{e(i)},
$$ 
let $F_i$ and $W_i$ be
shorthands for the sets $f_i(w_{i,1}, \ldots , w_{i,b+c-d+1})$ and
$\{w_{i,1}, \ldots ,w_{i,b+c-d+1}\}$, respectively, and let
$w_{i,j_1}, \ldots ,w_{i,j_{d-1}}$ be the first $d-1$ strings in $F_i$
(if $\|F_i\| \geq d$).  W.l.o.g., assume $F_i \seq W_i$ and $\|F_i\|
\leq c$ (if not, $f_i$ automatically is disqualified from being a
$\mbox{\rm GC}(b,c,d)$-selector).  

Define
$$
B_i \equalsdef \left\{
\begin{array}{ll}
\{ w_{i,j_1}, \ldots ,w_{i,j_{d-1}} \} \cup (W_i - F_i) &
\mbox{if $d \leq \|F_i\|$} \\
W_i & 
\mbox{if $d > \|F_i\|$.}
\end{array}
\right .
$$
Thus, either we have 
$$
\| W_i \cap
B\| \geq (d-1) + ((b+c-d+1) - c) = b \mbox{ and } \|F_i \cap B\| < d,
$$ 
or we have 
$$
\| W_i \cap B\| = b+c-d+1 > b \mbox{ and } \|F_i \cap B\| < d.
$$  
Hence, $B\not\in \mbox{\rm GC}(b,c,d)$.~\hfill$\Box$

\medskip

\noindent
{\bf Proof of Lemma~\ref{lemma:gc:inclusion}.}
\label{proof:gc:inclusion}
\ref{gc:inclusion:1}.\ \& \ref{gc:inclusion:2}. 
Immediate from the definitions of GC and S classes.  

\smallskip

\ref{gc:inclusion:3}.  Let $l \geq n$ and $m \geq n$. By
Parts~\ref{gc:inclusion:1} and \ref{gc:inclusion:2} of this lemma and
by Theorem~\ref{thm:S-collapse}, we have 
\begin{eqnarray*}
\mbox{\rm GC}(b,c,c) & = & \mbox{\rm S}(b,c) \ = \ \mbox{\rm S}(b+n, c+n) \ = \
                       \mbox{\rm GC}(b+n,c+n,c+n) \\
                 & \seq & \mbox{\rm GC}(b+l,c+m,c+n).
\end{eqnarray*}
 
\smallskip

\ref{gc:inclusion:4}. 
Suppose $m\leq l\leq n$ and $L\in \mbox{\rm GC}(b+l,c+m,d+n)$ via $f\in \fp$. 
As in the proof of Theorem~\ref{thm:S-collapse}, let finitely many strings 
$z_1, \ldots , z_{b+2l-1}$, each belonging to~$L$, be hard-coded into the
transducer computing function $g$ defined below. 
Given inputs $Y = \{y_1,  \ldots , y_t\}$, choose (if possible) $l$ strings
$z_{i_1},  \ldots , z_{i_l} \not\in Y$, 
and define
\[
g(Y) \equalsdef \left\{
\begin{array}{ll}
f(Y \cup \{z_{i_1}, \ldots , z_{i_l}\}) -
\{u{_1}, \ldots , u{_l}\} & 
\mbox{if $z_{i_1},  \ldots , z_{i_l} \not\in Y$ exist} \\
f(Y) - \{v{_1}, \ldots , v{_m}\} & \mbox{otherwise,}
\end{array}\right.
\]
where $\{u{_1}, \ldots , u{_l}\}$ contains {\em all\/} $z$-strings
output by $f$, say there are $h$ with $h \leq l$, the remaining $l-h$
$u$-strings are arbitrary $y$-strings of the output of $f$, and
similarly, $v{_1}, \ldots , v{_m}$ are arbitrary output strings of
$f$.  Clearly, $g \in \fp$ and $g(Y) \seq Y$.  Moreover, $\|g(Y)\|
\leq c+m-l \leq c$ if $z_{i_1}, \ldots , z_{i_l} \not\in Y$ exist;
otherwise, we trivially have $\|g(Y)\| \leq c$.  Note that if
$z_{i_1}, \ldots , z_{i_l} \not\in Y$ do not exist, then 
\[
\|Y \cap \{z_1, \ldots , z_{b+2l-1}\}\| \geq b+l .
\]  
Thus, if $\|L \cap Y\| \geq
b$, then either $\|L \cap (Y \cup \{z_{i_1}, \ldots , z_{i_l}\})\|
\geq b+l$ implies 
\[
\|L \cap g(Y)\| \geq d+n-l \geq d,
\] 
or $\|L \cap Y\| \geq b+l$ implies 
\[
\|L \cap g(Y)\| \geq d+n-m \geq d.
\]  
This establishes that $m\leq l\leq n$ implies 
\[
\mbox{\rm GC}(b+l,c+m,d+n) \seq
\mbox{\rm GC}(b,c,d).
\] 
By symmetry, we similarly obtain that $l\leq m\leq n$ implies the
containment of \mbox{$\mbox{\rm GC}(b+l,c+m,d+n)$} in \mbox{$\mbox{\rm
    GC}(b,c,d)$}, if we exchange $l$ and $m$ in the above argument.
  Since ($m\leq l\leq n$ or $l\leq m\leq n$) if and only if ($l\leq n$
  and $m\leq n$), the proof is complete.~\hfill$\Box$

\medskip

\noindent {\bf Proof of Lemma~\ref{lemma:gc:diag}.}
\label{proof:gc:diag}
\ref{gc:diag:1}.  
The diagonalization part of the proof is analogous to the proof of
Lemma~\ref{lemma:gc:trivial}.\ref{lem:gc:trivial:2}, the only
difference being that here we have $c+q$ instead of $c$.  Also, it
will be useful to require that any (potential) selector $f_i$ for some
set in $\mbox{\rm GC}(b,c+q,d)$ has the property that for any set of
inputs $W$ with $\|W\| \geq c+q$, $\|f_i(W)\|$ is {\em exactly\/}
$c+q$. By the remark after Definition~\ref{def:gch}, 
this results in an equivalent
definition of the GC class and can w.l.o.g.\ be assumed.  The
construction of set $L = \bigcup_{i\geq 1} L_i$ is as follows.  For
each $i$ with 
$$
2^{e(i)} < b+c+q-d+1,
$$ 
set $L_i \equalsdef \emptyset$.  For
each $i$ such that 
$$
2^{e(i)} \geq b+c-d+1,$$ 
let $F_i$ and $W_i$ be
shorthands for the sets $f_i(w_{i,1}, \ldots , w_{i,b+c+q-d+1})$ and
$\{w_{i,1}, \ldots ,w_{i,b+c+q-d+1}\}$, respectively, and let
$w_{i,j_1}, \ldots ,w_{i,j_{d-1}}$ be the first $d-1$ strings in $F_i$
(if $\|F_i\| \geq d$).  

If $\|F_i\| = c+q \, (\geq d)$ and $F_i \seq
W_i$, then set 
$$
L_i \equalsdef \{ w_{i,j_1}, \ldots ,w_{i,j_{d-1}} \} \cup (W_i - F_i);
$$ 
otherwise,
set $L_i \equalsdef W_i$. By the argument given in the proof of
Lemma~\ref{lemma:gc:trivial}.\ref{lem:gc:trivial:2}, 
$L \not\in \mbox{\rm GC}(b,c+q,d)$.

Now we prove that $L \in \mbox{\rm GC}(b+l,c+m,d+n)$ if $l>n$.  Given any
distinct input strings $y_1, \ldots ,y_t$, suppose they are
lexicographically ordered (i.e., \mbox{$y_1 <_{\mbox{\protect\scriptsize lex}}
\cdots <_{\mbox{\protect\scriptsize lex}} y_t$}), each $y_s$ is in $W_j$ for
some~$j$, and \mbox{$y_k <_{\mbox{\protect\scriptsize lex}} \cdots
<_{\mbox{\protect\scriptsize lex}} y_t$} are all strings of maximum length for
some $k$ with $1 \leq k \leq t$. Define a
$\mbox{\rm GC}(b+l,c+m,d+n)$-selector $f$ for $L$ as follows:
\begin{enumerate}
\item For $i \in \{1, \ldots , k-1\}$, decide by brute force whether
  $y_i$ is in $L$.  Let $v$ denote \mbox{$\|\{y_1, \ldots ,y_{k-1}\} \cap
  L\|$}. Output $\min\{v,d+n\}$ strings in $L$. If $v \geq d+n$ then
  halt, otherwise go to 2.

\item If $t\ge k+(d+n-v)-1$, then output $y_k, \ldots ,
  y_{k+(d+n-v)-1}$; otherwise, output $y_1, \ldots , y_t$.
\end{enumerate}
Clearly, $f \in \fp$, $f(y_1, \ldots ,y_t) \subseteq \{y_1, \ldots
,y_t\}$, and since $\mbox{\rm GC}(b+l,c+m,d+n)$ is nontrivial, we have:
$$\|f(y_1, \ldots ,y_t)\| \leq v+(d+n-v) \leq c+m.$$  

Now we prove that 
$$
\|\{y_1, \ldots ,y_t\}\cap L\|\ge b+l \Lora
\|f(y_1, \ldots ,y_t)\cap L\|\ge d+n.
$$  
Let $i$ be such that $e(i)$
is the length of $y_k, \ldots ,y_t$. Clearly, if $\|F_i\| \neq c+q$,
then by construction of $L$ and~$f$, 
either $f$ outputs $d+n$ strings
in~$L$, or 
$$
L \cap \{y_1, \ldots ,y_t\} = f(y_1, \ldots ,y_t),
$$
and so we are done.
Similarly, if $f$ halts in Step~1 because of $v \geq d+n$, then we are
also done.  

So suppose $v< d+n$, $\|\{y_1, \ldots ,y_t\}\cap L\|\geq b+l$,
and \mbox{$\|F_i\| = c+q \geq d$}.  Recall that $w_{i,j_{d-1}}$ is the
$(d-1)$st string in $F_i$.  Define 
\[
D \equalsdef \{y_k, \ldots ,y_t\} \cap
\{w_{i,1}, \ldots ,w_{i,j_{d-1}}\}.
\]  
By construction of $L$, we have
$ \{w_{i,1}, \ldots ,w_{i,j_{d-1}}\} \seq L$, so $D \seq L$. That is,
\begin{eqnarray}
\label{equ:diag-1}
\{y_k, \ldots ,y_{k+\|D\|-1}\} & \seq & L.
\end{eqnarray}
Since $\|\{y_k, \ldots , y_t\} \cap L\| \geq b+l-v$, we have
$$
t-(k-1) \geq b+l-v \geq d+n-v,$$ 
and thus, 
$$
t \geq k+(d+n-v)-1.
$$
This implies:
\begin{eqnarray}
\label{equ:diag-2}
\{y_k, \ldots ,y_{k+(d+n-v)-1}\} & \seq & f(y_1, \ldots ,y_t).
\end{eqnarray}
Thus, if $d+n-v \leq \|D\|$, we obtain from (\ref{equ:diag-1}) that
$\{y_k, \ldots ,y_{k+(d+n-v)-1}\} \seq L$, which in turn implies with
(\ref{equ:diag-2}) that 
$$
\|L \cap f(y_1, \ldots ,y_t)\| \geq v+(d+n-v)
= d+n.
$$ 

So it remains to show that $d+n-v \leq \|D\|$. Observe that
$$
b+l \leq \|\{y_1, \ldots ,y_t\}\cap L\| \leq v+\|D\|+b-d+1,$$ 
since
$\|W_i - F_i\| = (b+c+q-d+1) - (c+q) = b-d+1$ (here we need that
$\|F_i\| = c+q$ rather than $\|F_i\| \leq c+q$ for $f_i$ to be a
$\mbox{\rm GC}(b,c+q,d)$-selector).  Thus, $v+\|D\|+b-d+1 \geq b+l$. By
the assumption that $l\geq n+1$, we obtain $d+n-v \leq \| D\|$.

\smallskip

Parts~\ref{gc:diag:2} and~\ref{gc:diag:3} of this theorem can be
proven by similar arguments.~\hfill$\Box$

\bigskip

\noindent 
{\bf Acknowledgments}

We thank Hans-J\"org Burtschick, Johannes K\"obler, and Arfst
Nickelsen for interesting discussions. In particular, we are indebted
to Arfst Nickelsen for his kind permission to include his proof of the
first item in Theorem~\ref{thm:nick} and to Johannes K\"obler for
pointing out to us an error in an earlier version of this paper.

The first author was supported in part by grants NSF-CCR-8957604,
NSF-INT-9116781/JSPS-ENG-207,
NSF-INT-9513368/DAAD-315-PRO-fo-ab, and 
NSF-CCR-9322513.  
The second author was supported
in part by a postdoctoral fellowship from the Chinese Academy of
Sciences, and by grant NSF-CCR-8957604.  
The third author was
supported in part by a DAAD research visit grant, and grants
NSF-CCR-9322513, NSF-CCR-8957604, and
NSF-INT-9513368/DAAD-315-PRO-fo-ab.  
The fourth author was supported
in part by grant NSF-INT-9116781/JSPS-ENG-207.

This work was done in part while the first author was visiting
Friedrich-Schiller-Universit\"at Jena and the Tokyo Institute of
Technology, and in part while the other three authors were visiting 
the University of Rochester. 

\clearpage

\bibliographystyle{alpha}

{\singlespacing 
\bibliography{gry}
}

\end{document}